\documentclass[prl,aps,reprint,notitlepage]{revtex4-2} 
\usepackage{xcolor}
\usepackage{blindtext}
\usepackage{titlesec}
\usepackage{amsmath,amssymb,bbold,bm}
\usepackage{graphicx}
\usepackage{cancel}
\usepackage{natbib,hyperref}

\newcommand{\sbraket}[1]{\langle #1 \rangle}

\newcommand{\be}{\begin{equation}}
\newcommand{\ben}{\begin{equation*}}
\newcommand{\ee}{\end{equation}}
\newcommand{\een}{\end{equation*}}
\newcommand{\bs}{\begin{split}}
\newcommand{\es}{\end{split}}
\newcommand{\bmx}{\begin{array}}
\newcommand{\emx}{\end{array}}

\newcommand{\bea}{\begin{eqnarray}}
\newcommand{\bean}{\begin{eqnarray*}}
\newcommand{\eea}{\end{eqnarray}}
\newcommand{\eean}{\end{eqnarray*}}
\newcommand{\dg}{^{\dagger}}

\newcommand{\lr}{\leftrightarrow}

\newcommand{\bb}[1]{\mathbb{#1}}
\newcommand{\qqquad}{\qquad\qquad\qquad}
\newcommand{\so}{\qquad\rightarrow\qquad}

\newcommand{\orr}{\qquad\text{or}\qquad}
\newcommand{\andd}{\qquad\text{and}\qquad}

\newcommand{\eps}{\epsilon}

\newcommand{\pref}[1]{(\ref{#1})}

\newcommand{\tr}[1]{{\rm Tr}[#1]}

\newcommand{\abs}[1]{\left\vert #1 \right\vert}

\newcommand{\braket}[1]{\left\langle #1\right\rangle}

\newcommand{\mat}[1]{\left(\bmx{cc}#1\emx\right)}

\newcommand{\matn}[1]{\bmx{cc}#1\emx}
\newcommand{\matl}[1]{\bmx{ll}#1\emx}

\newcommand{\sepline}{}

\newcommand{\bw}[1]{\begin{widetext}}
\newcommand{\ew}[1]{\end{widetext}}

\newcommand{\gray}[1]{}

\newcommand{\nothing}[1]{}

\begin{document}

\title{Higher-Dimensional Chirally Stabilized Fixed Points and Their Deformations}
\author{Aleksandar Ljepoja}
\author{L.\,C.\,R. Wijewardhana}
\author{Yashar Komijani\,$^{*}$}
 \affiliation{ Department of Physics, University of Cincinnati, Cincinnati, Ohio, 45221, USA}
\date{\today}
\begin{abstract}
Non-Fermi liquids in $d>2$ remain poorly understood, particularly when relevant perturbations destabilize them. In one spatial dimension, chirally stabilized fixed points provide a rare class of analytically tractable non-Fermi-liquid critical points, but their higher-dimensional analogues have been elusive. Here, we develop a Wilsonian operator–product–expansion renormalization group scheme that captures power-divergent terms and use it to construct finite-$N$ higher-dimensional analogues of chirally stabilized fixed points in arbitrary dimension $d\le4$. This exposes a conformal window at finite $N$. We further show that symmetry-breaking masses, far from being trivial, can collapse this window and drive the system to strong coupling, triggering dynamical mass generation.
\end{abstract}
\maketitle

\emph{Introduction} -- 
Classification of non-Fermi liquid systems and their fate upon destabilizing perturbations is one of the major open questions in condensed matter physics, particularly in higher dimensions, where the powerful toolkit of 1+1D methods is inapplicable. 

In one spatial dimension, the integrability and conformal field theory (CFT) provide exact control over certain non-Fermi-liquid fixed points. A particular example is the chirally stabilized fixed points, a family of critical infrared (IR) fixed points in 1+1D which have played a central role in understanding impurity models, quantum wires, and topological phases. They result from $J\bar J$ deformations of Wess-Zumino-Witten CFT, where $J$ and $\bar J$ are left- and right-moving currents in some affine Lie algebra, e.g. SU($N$)$_K$. When the currents have the same affine levels, the theory is believed to flow to a gapped ground state. However, in the level-asymmetric case \cite{SOLOVIEV1996}, the IR fixed point is expected to be governed by a non-trivial critical fixed point. In the case of SU(2)$_K$ and $\overline{\rm SU(2)}_1$,  by matching conformal and current-level anomalies \cite{Andrei1998,Andrei2000} the IR fixed point is conjectured to be the SU(2)$_{K-1}\otimes\overline{\cal M}_{K+1}$ theory where ${\cal M}_{K+1}$ is the coset model equivalence of the minimal series of CFT, a prediction that is corroborated by Bethe ansatz \cite{Andrei1998}. The $K=2$ case has been studied explicitly at the so-called Toulouse point \cite{Azaria1998,Azaria2000,Azaria2000b} using abelian bosonization, and has also been utilized recently to study the topological order in 2+1D within the so-called coupled-wire construction \cite{Ljepoja2024}. This sets the stage for the question: can analogous fixed points exist in higher dimensions, and if so, how can they be controlled?

In higher dimensions, such fixed points are far less tractable: the loss of exact 1+1D methods and the limitations of large-N or dimensional regularization schemes leave open fundamental questions about their existence, stability, and possible deformations. One tractable case is the $N,K\to\infty$ limit while $\kappa=K/N$ is kept finite. In this \emph{dynamical large-$N$} limit, new conformal IR fixed points were found in 1+1D \cite{Ge2022} and 2+1D \cite{Ge2024} in the study of the multi-channel Kondo lattices, with the latter suggested as the 3D analogue of Banks-Zaks fixed point \cite{Banks1982}. While in these works the Lorentz symmetry was broken by the interaction in the microscopic lattice model, the re-emergence of the symmetry at intermediate temperatures \cite{Ge2022,Ge2024}, suggests that a relativistic model provides a minimal and faithful description of the ultraviolet (UV) physics, enabling us to verify these lattice results and see if they persist down to finite $N$.
However, controlling interacting fixed points in $d>2$ with finite $N$ remains formidable. Related four-fermion models, such as the Gross–Neveu and Thirring models, have been studied extensively in $d>2$ using large-$N$, functional RG, and lattice methods \cite{Semenoff:1986dv,Semenoff:1989dm,Rosenstein1991,Bennett1999,Hofling2002,Appelquist:2004ib,Gies2010,Janssen2012,Hands2019,Hands2020b,Wipf2022,AriasTamargo2022} but these works focused on flavor-symmetric single-species interactions and left finite-$N$ questions unsettled. 

This gap motivates us to study the generalization of such a model to arbitrary dimensions, and study a \emph{flavor-asymmetric} \emph{two-species} non-abelian Thirring model in spacetime dimensions  $d\le 4$ and finite $N$. Within the renormalization group (RG) framework, we find that the IR behavior of the model is tractable for arbitrary $\kappa$ at $d=2+\eps$. We show that for a wide range of parameters, a conformal window \cite{Appelquist:1988sr} exists in which a sufficiently large coupling flows to a non-trivial finite coupling fixed point. We compute the critical exponents at this fixed point and compare our results to dynamical large-N. 

Next we analyze deformation of the theory by adding a symmetry-breaking perturbation term \cite{Semenoff2025} to one of the species. We use the Wilsonian momentum-cutoff scheme, rather than dimensional regularization \cite{Bennett1999,Hofling2002,Gies2010,Hands2019,Hands2020b,Wipf2022,AriasTamargo2022}, which enables us to study the non-perturbative effects of such massive deformations. While the interactions become irrelevant in the long distances compared to the induced gap, we find that mass deformations, far from being trivial, alters the intermediate-distance flow and shrinks the conformal window and in some cases triggers dynamical mass generation.
We discuss the manifold of the order parameter at such strong coupling fixed points.

\emph{The model} -- To address these questions in a controlled setting, we introduce the following two-species model. The Euclidean action in $d$ spacetime dimensions is $S=\int{d^dr}{\cal L}$ where using the implicit summation over repeated indices, $\bar\psi\equiv\psi\dg\gamma^0$ and $\bar\chi\equiv\chi\dg\gamma^0$,
\be
{\cal L}=\bar\psi_{\alpha a}(\cancel\partial+m_0\hat m)\psi_{\alpha a}+\bar\chi_{\alpha b}\cancel\partial\chi_{\alpha b}+\frac{\lambda_0}{N}{\cal J}^{A,\mu}_\psi{\cal J}_{A,\mu}^\chi.\label{eq1}
\ee
Here, $\psi_{\alpha a}$ and $\chi_{\alpha b}$ are two sets of Dirac fermions, with $\alpha=1\dots N$ for the common spin (color) index, whereas $a=1\dots K_\psi$ and $b=1\dots K_\chi$ are the channel (flavor) indices. We assume $K_\psi\ge K_\chi$. $\cancel\partial\equiv\gamma^\mu\partial_\mu$ with $\mu=0\dots d-1$. We are interested in $d=2,3, 4$ and therefore choose $\{\gamma^\mu,\gamma^\nu\}=2\delta^{\mu\nu}\bb 1_{4\times4}$ with the normalization ${\rm Tr}[{\gamma^\mu\gamma^\nu}]=4\delta^{\mu\nu}$. Note that a hybridization $V(\bar\psi\chi+\bar\chi\psi)$ could open a gap in the spectrum in the flavor-symmetric case \cite{Ftnote2}.
The currents ${\cal J}^{A,\mu}_\psi\equiv\bar\psi_{a\alpha} T^A_{\alpha\beta}\gamma^\mu\psi_{a\beta}$ and ${\cal J}^{A,\mu}_\chi\equiv\bar\chi_{b\alpha} T^A_{\alpha\beta}\gamma^\mu\chi_{b\beta}$ are defined in terms of $T^A$ with $A=1\dots N^2-1$ which belong to the fundamental representation of SU($N$) and act in the color space. They are normalized such that $\tr{T^A}=0$, $\tr{T^AT^B}=\delta^{AB}$,  $[T^A,T^B]=if^{ABC}T^C$ and $\{T^A,T^B\}={\cal D}^{AB}$. In addition to the diagonal SU($N$) symmetry, the UV theory has U($K_\psi$)$\times $U($K_\chi$) flavor-charge symmetry. 

The vector model \pref{eq1} is equivalent to the strong coupling, or Nambu-Jona-Lasinio (NJL), limit of a two-species QCD model where $\psi$ and $\chi$ interact via two sets of non-abelian gauge fields. Although the interaction is non-chiral, in $d=2$  it decouples to ${\cal J}_\psi \bar {\cal J}_\chi+{\cal J}_\chi \bar {\cal J}_\psi$ parts, each flowing to a chirally stabilized fixed point \cite{Andrei1998}. In $d=3$, it is the simplest two-species fermion model on a honeycomb lattice \cite{SM} with an interaction transforming as an adjoint representation of SU($N$). In this case, the $\gamma_{4\times 4}$ matrices are reducible, and they can be chosen to be $\gamma^\mu=\{\sigma^z,\sigma^y,-\sigma^x\}\otimes\tau^0$ where $\sigma^\mu$ and $\tau^\nu$ are two sets of Pauli matrices acting on sublattice and valley, respectively. The mass term in \pref{eq1} can be any matrix that satisfies $[\hat m,\gamma^\mu]=0$. With our choice of $\gamma^\mu$, the four choices include the time-reversal symmetry breaking $\hat m=\bb 1$, transforming as a singlet under pseudospin-valley (PSV) symmetry, as well as
parity-symmetry breaking $\tau^z$ and Kekul\'e order parameters $\tau^x$ and $\tau^y$ which transform as a triplet. In $d=4$ the model \pref{eq1} corresponds to interacting Weyl semimetals.

The Dirac fermions have the engineering dimension $\Delta_0=[\psi,\chi]=(d-1)/2$, so that $[\lambda]=2-d$ and the interaction is irrelevant for $d>2$, whereas the mass term always relevant, with $[m]=1$. We define dimensionless couplings $\lambda(a)\equiv\lambda_0a^{2-d}$ and $m(a)\equiv m_0a$. We first study the model in the massless limit $m=0$ and then consider the $m\neq 0$ deformations.

\begin{figure}[tp!]
\includegraphics[width=\linewidth]{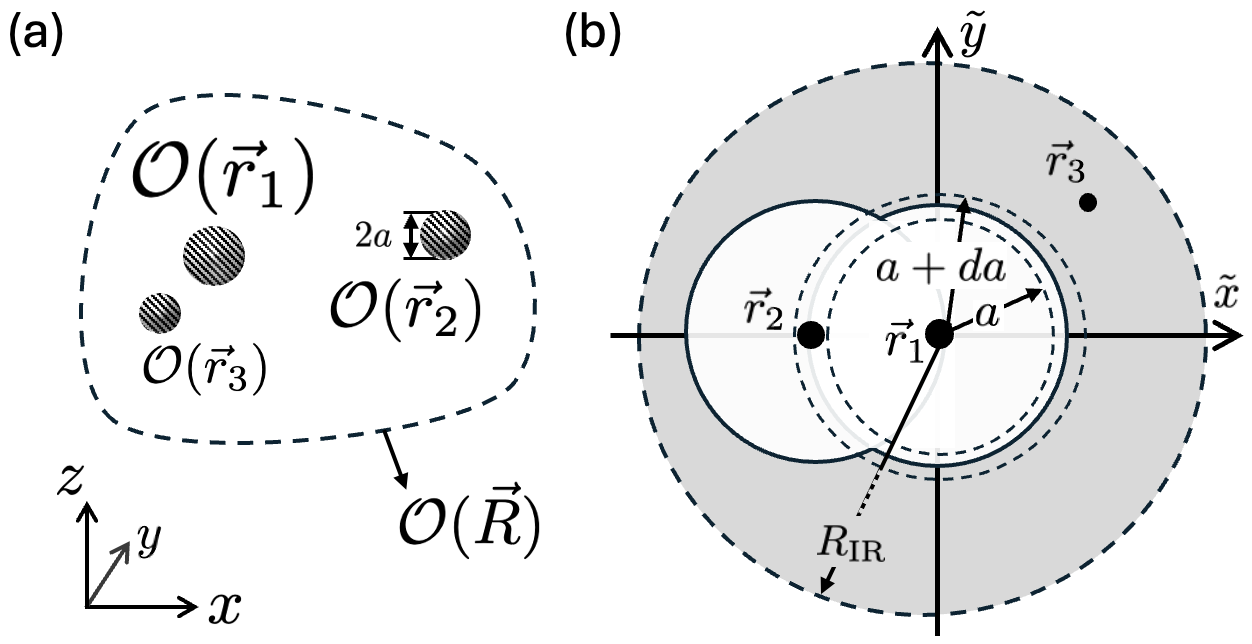}
\caption{\small (a) Multiple insertions of the interaction ${\cal O}(\vec r)={\cal J}_\psi^{\mu A}{\cal J}^\chi_{\mu A}$ in spacetime $\vec r_j$  coalesce into ${\cal O}(\vec R)$ in a center of mass position. The points cannot get closer than the small scale cutoff $a$. (b) The two-loop shell integration involves a free rotation of $\vec r_2$ and an effectively 2D integral over $\vec r_3$.}\label{Fig1}\vspace{-.4cm}
\end{figure}

\emph{RG Analysis} --  To see how the couplings evolve as the bandwidth $\Lambda$ is reduced, we do a two-loop Wilsonian RG analysis in $d$ dimensions in the so-called Cardy's scheme \cite{Cardy2015}. We consider insertion of interactions at space-time points $\vec r_i$, and demand that these points are not closer to each other than the small-distance cut off $a=1/\Lambda$, thus representing each with a solid ball [Fig.\,\pref{Fig1}(a)]. Reducing momentum cutoff $\Lambda$ corresponds to increasing $a$ by integrating out all the insertions within a shell $(a,a+da)$. 
 
\begin{figure}[tp!]
\includegraphics[width=\linewidth]{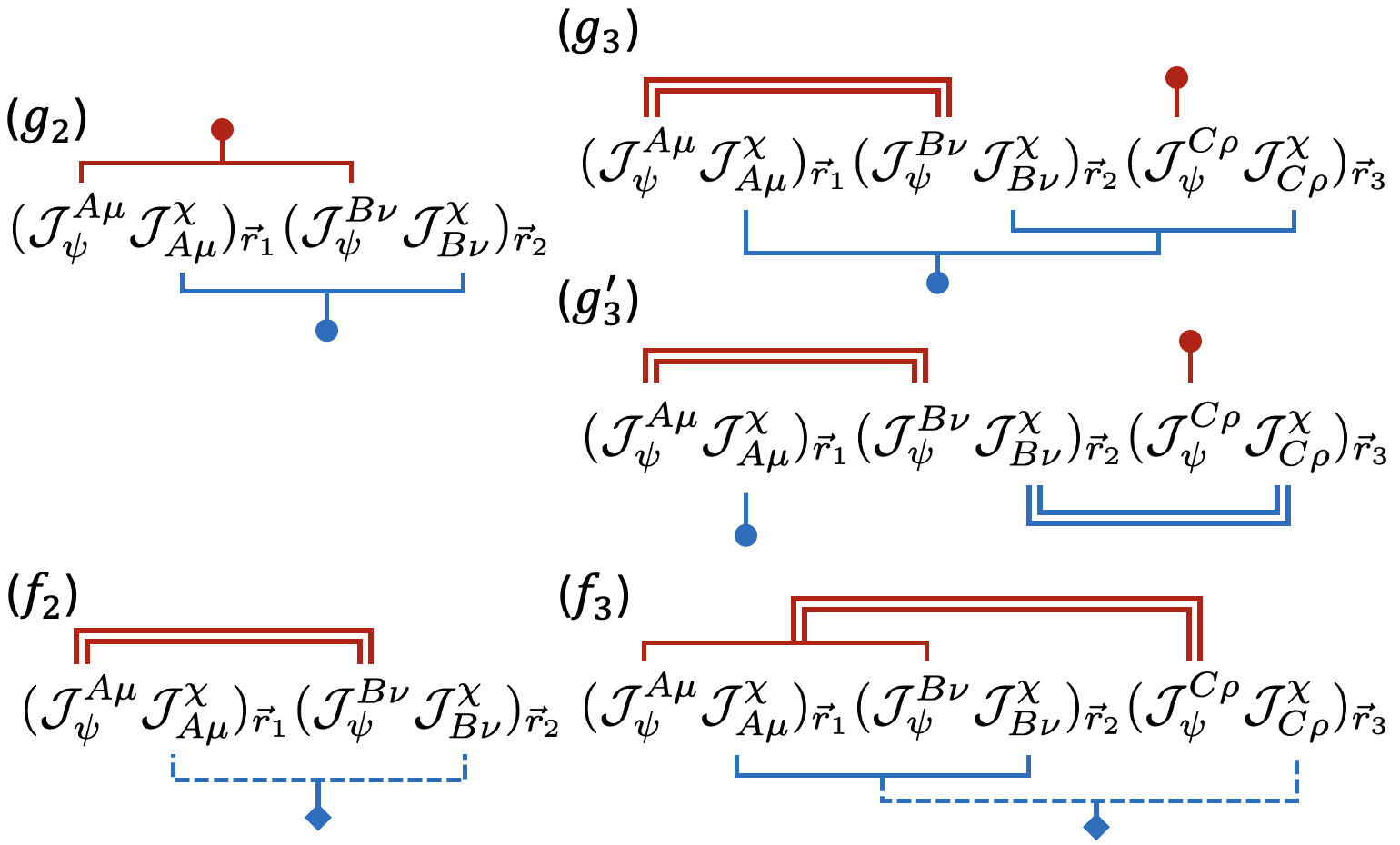}
\caption{Diagrams contributing to the renormalization of ($g_i$) interaction vertex and ($f_i$) wavefunction renormalization, in one-loop $\propto\lambda^2$ and two-loop $\propto\lambda^3$ orders, up to trivial permutations. The single- and double-lines correspond to single- and double-contractions in the operator product expansions of Eq.\,\pref{ope}. The circle (solid line) and diamond (dashed line) denote ${\cal J}$-currents and $\bar\chi\cancel\partial\chi$, respectively.}
\vspace{-.2cm}\label{Fig2}
\end{figure}

The latter is done using the so-called operator product expansion (OPE), which describes how the fields in the theory fuse to generate new operators in the long distance [Fig.\,\ref{Fig1}(a)]. Using $\sbraket{\psi(\vec r)\bar\psi}= \cancel r/\Omega_d r^d$, expressed in terms of $r=\abs{\vec r}$ and the solid angle $\Omega_d=2\pi^{d/2}/\Gamma(d/2)$, the required OPE is \cite{SM}
\bea
&&{\cal J}_\psi^{A,\mu}(\vec r){\cal J}_\psi^{B,\nu}\sim K_\psi\delta^{AB}\frac{4}{\Omega_d^{2}r^{2d}}C^{\mu\nu}(\vec r)\label{ope}\\
&&\hspace{1cm}+\frac{1}{\Omega_dr^d}B^{\mu\nu}_\lambda(\vec r)[if^{ABC}{\cal J}_\psi^{C,\lambda}-\bar\psi{\cal D}^{AB}\gamma^\lambda\partial_b\psi]+\dots\nonumber
\eea
where $C^{\mu\nu}(\vec r)=2r^\mu r^\nu-r^2\delta^{\mu\nu}$ and $B^{\mu\nu}_\lambda=\frac{1}{2}\partial C^{\mu\nu}/\partial r^\lambda$. The same holds for $\chi$ fermions. This OPE is valid in $d=2$ and $d=4$ as well as the continuous $2\le d\le 4$ as long as ${\rm Tr}[{\gamma^\mu\gamma^\nu\gamma^\rho}]=0$, which we assume to hold in the spirit of $d=2+\eps$ expansion \cite{Ftnote1}. The OPE \pref{ope} means that inside any correlation function, the expression on the left can be replaced by the expression on the right, as long as other operators are sufficiently far from these points. The ``$\dots$'' indicate less relevant terms, which are also non-singular in $d\le 4$ and vanish if the operators on the left approach each other. We will refer to the first term ($\propto K$) and the second term in \pref{ope} as double- and single-contractions, and represent them by double- and single-lines in the diagrams [Fig.\,\ref{Fig2}], respectively.

Multiple vertex insertions coalesce [Fig.\,\ref{Fig1}(a)] via OPEs into various terms of the Lagrangian at the center of mass \cite{Affleck1991}.  A 2D plane can be passed through any three spacetime insertions, reducing the integrals to two-dimensional ones. Furthermore, by placing $\vec r_1$ at the center of coordinate and a subsequent rotation, the three points can be brought to the form depicted in Fig.\,\ref{Fig1}(b). The points cannot be closer than the short-distance cutoff $a$. Furthermore, to contribute to the RG flow, one and only one point has to be inside the shell $(a,a+da)$. Therefore, one-loop calculations reduce to a trivial rotation integral and two-loop calculations involve a non-trivial two-dimensional integrals, covering the grey area in Fig.\,\ref{Fig1}(b). These integrals are computed semi-analytically \cite{SM,Gamayun2023}. 

The irreducible contractions contributing to the interaction vertex and wavefunction renormalizations are sketched in Fig.\,\ref{Fig2}($g_i$) and Fig.\,\ref{Fig2}($f_i$), respectively, up to trivial permutations. Any contraction not shown (for example purely single-contractions) is either reducible or zero \cite{Ludwig2003}. In particular, the new diagram \ref{Fig2}($g_{3}'$) with the prefactor $\kappa_\psi\kappa_\chi$ vanishes for $d=2$ and is discarded in dimensional regularization scheme. However, it plays an important role in our discussion of deformations.

Integrating out the shell modifies the Lagrangian to ${\cal L}\to Z_\psi\bar\psi\cancel\partial\psi+Z_\chi\bar\chi\cancel\partial\chi+(\lambda_0/N)Z_\lambda{\cal J}_\psi^{A\mu}{\cal J}^\chi_{A\mu}$, where $Z_X=1+y_Xd\ell$ for $X=\lambda, \psi, \chi$ are corrections with $d\ell\equiv da/a$. Rescaling the spacetime $\vec r\to (1+d\ell)\vec r$ and fields $\psi\to (1+d\ell)^{(1-d)/2}Z_\psi^{-1/2}\psi$ (similarly for $\chi$) returns the kinetic terms to their original form. The anomalous dimension of the fields are then obtained from $\delta_\psi=\frac{1}{2}y_\psi$ and $\delta_\chi=\frac{1}{2}y_\chi$, while the beta function is found from $Z_\lambda/Z_\chi Z_\psi$.

\emph{Results} -- For the $\chi$-fields we have $\Delta_\chi=(d-1)/2+\delta_\chi$ where the
anomalous dimension is given by
\be
\delta_\chi=\frac{1}{2}\kappa_\psi\big[f_2\lambda^2+f_3\lambda^3\big]+O(\kappa_\psi\lambda^4),
\ee
with a $\psi\lr\chi$ substitution for  $\delta_\psi$. We have pulled out the re-scaled flavor numbers $\kappa_X\equiv K_X/N$ for $X=\psi,\chi$. Besides this re-scaling, the $N$-dependence is rather weak. The beta function is
\be
\frac{d\lambda}{d\ell}=\beta(\lambda)=(2-d)\lambda+g_2\lambda^2-b\lambda^3+O(\kappa_\psi \kappa_\chi\lambda^4),\label{eqbeta}
\ee
where $b=(\kappa_\chi+\kappa_\psi)(f_2+g_{3})-\kappa_\psi\kappa_\chi g_{3}'$. The coefficients $g_i(d)$  and $f_i(d,N)=(1-N^{-2})\tilde f_i(d)$, with their mild $N$-dependence, are positive and smooth functions of $d$, except $g'_{3}\propto (d-2)$ which vanishes at $d=2$. These $O(1)$ coefficients are presented in the end matter, but the discussion below is independent of their precise values.

\begin{figure}[tp!]
\includegraphics[width=\linewidth]{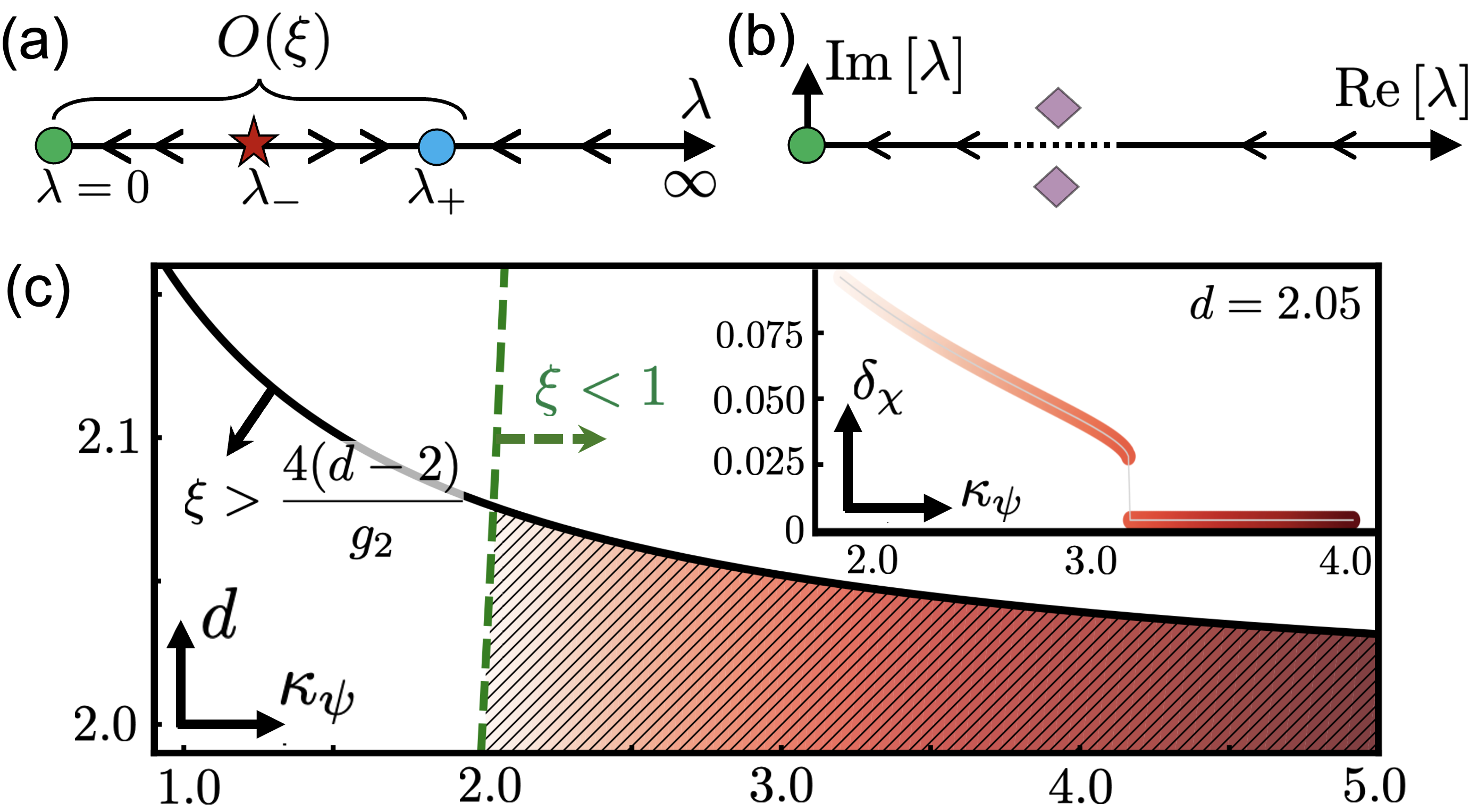}
\caption{\small The RG flow for $\lambda$ for parameters $\kappa_\psi,\kappa_\chi,d,N$ (a) inside conformal window and (b) outside conformal window. The fixed point $\lambda_0$, $\lambda_-$ and $\lambda_+$ denote the (stable) non-interacting, the (unstable) critical point and the (stable) non-trivial fixed point, respectively. For $d=2+\eps$, $\lambda_-$ is within the perturbative $\lambda\sim O(\eps)$ limit. At large-flavor limit $\kappa\to \kappa_c$, the $\lambda_\pm$ merge and eventually go off real axis, creating a walking RG flow of the coupling. (c) The reliable conformal window $\lambda_-<\xi< 1$ is marked in the red region as a function of $\kappa_\psi$ and $d$. (Inset) The anomalous dimension $\delta_\chi$ vs. $\kappa_\psi$ decays and jumps to zero at the loss of criticality.
}\label{Fig3}\vspace{-.5cm}
\end{figure}

In terms of $\xi\equiv g_2/b$, the fixed points $\beta(\lambda_*)=0$ are
\be
\lambda=0,\andd \lambda_\pm=\frac{\xi}{2}\Big[1\pm\sqrt{1-\frac{4(d-2)}{g_2\xi}}\Big].
\ee
 Using $d\lambda/d\ell \approx (\lambda - \lambda_*)\beta'(\lambda_*)$ near a fixed point, the sign of $\beta'(\lambda_*)$ reveals that when the roots are real, $\lambda_-$ is unstable, while both $\lambda=0$ and $\lambda_+$ are stable. 
 Although the interaction is irrelevant, if the coupling exceeds a threshold $\lambda > \lambda_-$, it becomes relevant and flows toward the non-trivial fixed point at $\lambda_+$ [Fig.\,\ref{Fig3}(a)]. For the marginal $d=2$ case, the threshold $\lambda_-$ vanishes and any coupling flows to $\lambda_+$. In the $d = 2$ and the flavor-symmetric case $\kappa_\psi = \kappa_\chi$, the beta function reduces to known results \cite{Bennett1999}.
 
 Remarkably, for $d>2$, the condition for real fixed points, $4b(d - 2) < g_2^2$, defines a \emph{conformal window} in the parameter space of $\kappa_\chi$, $\kappa_\psi$, $N$, and $d$. In this window, the model flows to the nontrivial fixed point. Outside this window, the fixed points $\lambda_\pm$ collide and move off the real axis [Fig.\,\ref{Fig3}(b)], and the interaction becomes irrelevant. Nevertheless, the RG flow exhibits a slow, \emph{walking} behavior near the complex conjugate roots of $\beta(\lambda)$ \cite{Kaplan2009,Gorbenko2018}.

Earlier studies entirely missed this conformal window by focusing on the flavor-symmetric model. Indeed, the flavor-symmetric  $d=2$ case of \pref{eqbeta} illustrates the limitations of perturbative RG \cite{Bennett1999,Gerganov2001,Itsios2014}: Taking the $\kappa_{\chi,\psi}\to\kappa$ limit, we see that the $O(\kappa^2\lambda^4)$ term in Eq.~\pref{eqbeta} drives the non-trivial fixed point $\lambda_+\propto 1/\kappa$ to strong coupling. Thus, the apparent persistence of the conformal fixed point at two-loop order is a \emph{mirage}  \cite{Kaplan2009,Gorbenko2018}, as revealed by the resummation of the beta function in the large-$\kappa$ limit \cite{Kutasov1989}. This is even more restrictive in $d>2$ due to the $g'_3$ term which implies that $\kappa_\chi>\kappa_c$, with $\kappa_c\equiv(1+\kappa_\chi/\kappa_\psi)(f_2+g_3)/g_3'$ changes then sign of the $b$ term leading to the strong coupling. Hereafter, we focus on the highly asymmetric regime $\kappa_\chi/\kappa_\psi\to 0$ and $\kappa_\chi<\kappa_c$ where a conformal window can be safely defined.

 The perturbative RG result is reliable as long as $\lambda_\pm\ll 1$, so that the fixed point are close to the Gaussian fixed point. A necessary condition is the dimensional continuation $d=2+\eps$, but while $\eps\to 0$ results in $\lambda_-\approx \eps/g_2$, the $\lambda_+\approx \xi$ remains finite. Therefore, we demand more generally that ${4(d-2)}/g_2<\xi\ll 1$, which defines a hyper-volume of fixed points. One special case, relevant to dynamical large-$N$ is when $\kappa_\chi\to 0$, but $\kappa_\psi$ is kept finite. Another special case, relevant to finite-$N$ and large channel limit, is when $\kappa_\psi\to\infty$. In both cases $\delta_\psi\to 0$, $\delta_\chi\sim \kappa_\psi^{-1}$ and $\xi=\kappa_\psi^{-1}g_2/[f_2+g_{3}-g'_{3}\kappa_\chi]$. Not relying on dimensional regularization pays off later when we depart the massless regime. 
 
Fig.\,\ref{Fig3}(c) depicts the $\xi(\kappa_\psi,d)$ for $N\to\infty$ and $\kappa_\chi=0$, highlighting that a conformal window exists for a wide range of parameters, which include $\xi< 1$ (red hashed region). The anomalous dimension $\delta_\chi(\kappa_\psi)\sim 1/\kappa_\psi$ decays until it jumps to zero at the edge of conformal window.

\emph{Deformation} -- Next we turn on the symmetry-breaking mass term $m_0\neq 0$, either along $\hat m=\bb 1$ or one of $\hat m=\vec\tau$. The goal is to see possible non-trivial effects before the bandwidth is reduced below the mass gap of $\psi$ fermions. The mass modifies the $\psi$ Green's function \cite{SM} to $\sbraket{\psi(\vec r)\bar\psi}= (\cancel r+nr\hat m)g/\Omega_d r^d$. The new parameters $g(x)=2^{1-d/2}x^{d/2}K_{d/2}(x)$ and $n(x)=K_{d/2-1}(x)/K_{d/2}(x)$ are dimensionless functions of $x=m_0r$, expressed in terms of Bessel functions. At small $x$, $g(x\ll 1)\to 1$ but at large $x$, $g(x\gg 1)\propto x^{-(d-1)/2}e^{-x}$. The latter essentially sets a new IR cut-off $R_{\rm IR}\to 1/m_0$ instead of the size of the system, an effect that is unimportant as long as $\Lambda\gg m_0$.  On the other hand $n(x)$ has the asymptotics $n(x\ll 1)\to x/(d-2)$ and $n(x\gg 1)\to 1$. For the OPE \pref{ope}, $B^{\mu\nu}_\lambda$ remains intact, but the mass-induced $nr\hat m$ term in the propagator modifies $C^{\mu\nu}\to 2r^\mu r^\nu-r^2(1+n^2)\delta^{\mu\nu}$. This change is independent of $\hat m$ and thus all massive deformations have the same effect \cite{SM}. In addition, the OPE gets a new term $r^{1-d}gn\bar\psi {\cal D}^{AB}\psi$ \cite{SM} which is responsible for mass renormalization $Z_m$, and the sub-leading correction to the mass beta function $dm/d\ell=m+O(\lambda^2)$. For brevity, we omit discussing these corrections and focus on the RG flow of $\lambda$ and the anomalous dimensions of the fields.

\begin{figure}[tp!]
\includegraphics[width=1\linewidth]{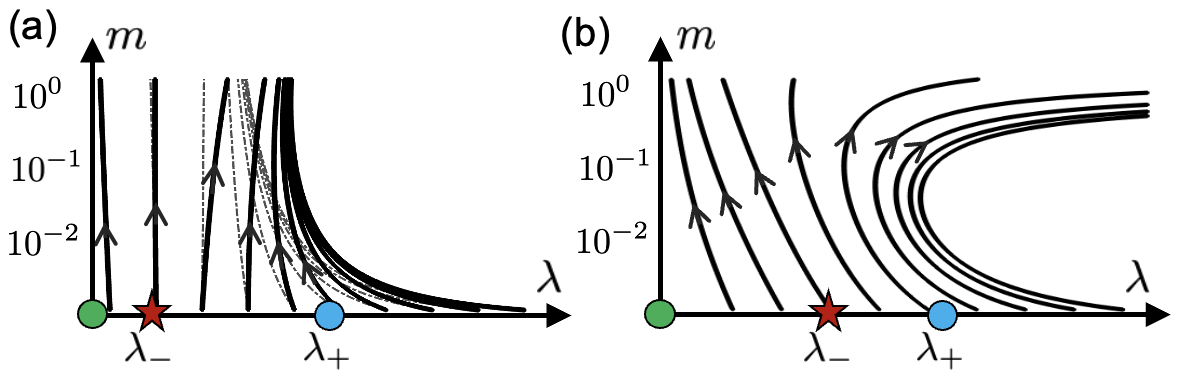}
\caption{\small\raggedright The RG flow on the $(\lambda,m)$ space. (a) For $\kappa_\chi<\kappa_c$ the symmetry breaking alters the flow (compared to the massless case, dashed grey) only slightly before reaching the mass gap $m\sim O(1)$, eventually leading to a trivial fixed point. (b) For $\kappa_c'<\kappa_\chi<\kappa_c$, the modification is substantial, causing $\lambda$ to flow to the strong coupling $O(1)$, and dynamic mass generation, before reaching the trivializing mass gap.}
\label{Fig4}\vspace{-.5cm}
\end{figure}

The one-loop diagrams $f_2$ and $g_2$ are not affected in this limit, as the distance between the spacetime insertions $m=m_0a\ll 1$, makes the $n(m)$ negligible. In contrast, in the two-loop contributions $f_3$, $g_3$ and $g'_3$, the free insertion ($\vec r_3$ in Fig.\,\ref{Fig1}a) is integrated from UV up to the $R_{\rm IR}\sim 1/m_0$, and  therefore  they are affected by the $n(m_0r)$ term. The main modification is to the $g_3'$ term in the form of the \emph{non-perturbative} correction
\be
g_3'\to g_3'+g_3''\abs{m}^{d-2}+O(m^2).\label{eqg3p}
\ee
Unlike $g'_3$ itself, the correction $g''_3(d=2+\eps)\sim 0.07$ does not vanish in $d\to 2$ \cite{SM}. One can regard the change $\Delta g_3'$ as a running coupling, with the beta function $d\Delta g_3'/d\ell=(d-2)\Delta g_3'$. There are also $O(m^2)$ corrections to the all other coefficients, but those corrections are negligible in the short distance limit where $m\ll 1$. Unlike these, the non-analytic contribution \pref{eqg3p} has a chance to compete parametrically with the two loop terms. Since $m=m_0/\Lambda$ is relevant, the nearly logarithmic correction \pref{eqg3p} causes a RG-persistent downward shift to the two-loop coefficient $b$ which shrinks the conformal window [Fig.\,\ref{Fig4}(a)]. The $\kappa_\chi<\kappa_c$ threshold for conformal fixed point is modified to $\kappa_c'\sim(1+\kappa_\chi/\kappa_\psi)(f_2+g_3)/(g_3'+g_3'')$. Since $g'_3(d=2)=0$ this represents a significant change in $d=2+\eps$.

The change in $g'_3$ introduces marginal changes to the flow until the bandwidth reaches the mass gap $m\sim O(1)$, after which the coupling becomes irrelevant. Most dramatically, for $\kappa_c'<\kappa_\chi<\kappa_c$ symmetry breaking drives $b\to 0$, sending $\lambda_+\to\infty$ \emph{before} the mass gap is reached [Fig.\,\ref{Fig4}(b)]. In this case, symmetry breaking causes dynamical mass generation as we discuss next. 

\emph{Strong coupling} -- Using Fierz identity \cite{SM} the interaction can be written as 
\be
{\cal L}_{int}=-\frac{\lambda_0}{N}\Big[(\bar\psi_{a}\tau^\nu\chi_{b})(\bar\chi_{b}\tau_\nu\psi_{a})+\frac{(\bar\psi\gamma^\mu\psi)(\bar\chi\gamma_\mu\chi)}{N}\Big],
\ee
where $\nu=0,x,y,z$, originating from an {accidental} O(4) PSV symmetry \cite{Ftnote3}. The second term is a 1/$N$ suppressed current-current interaction which is irrelevant in the RG sense. In the strong coupling, the first term leads to the formation of the spin/color-singlet boson $V^\nu_{ab}\sim{\bar\psi_{\alpha a}\tau^\nu\chi_{\alpha b}}$, which is a PSV-singlet and -triplet for $\nu=0$ and $\nu=x,y,z$, respectively. It hybridizes the two fermions and opens a gap in the spectrum. The bosonic matrix $V^\nu_{ab}$ transforms as a bi-fundamental representation of $U(K_\psi)\times U(K_\chi)$ and 
appears as a bound state in $\braket{V^\nu_{ab}(\vec r)\bar V^\nu_{ab}}$ correlation function \cite{Ge2024}. The condensation of $\braket{V^\nu_{ab}}$ spontaneously breaks the PSV symmetry $O(4)\to O(3)$ \cite{Ftnote3}. Additionally, a rank-$r$ $\braket{V^\nu_{ab}}$ breaks the flavor symmetry \cite{Ftnote4,Wugalter2020,Ge2024} and hybridizes $r$ flavors of $\psi$ and $\chi$ together, with an order parameter manifold
\be
\frac{G}{H}\cong \frac{U(K_\psi)\times U(K_\chi)}{U(K_\psi-r)\times U(K_\chi-r)\times U_{\rm diag}(r)}\times \frac{O(4)}{O(3)}.\label{eqGH}
\ee
The $U_{\rm diag}(r)$ mirrors the dynamical mass generation in 1+1D  Gross–Neveu model \cite{Witten1978} where a chiral U(1) symmetry is spontaneously broken, by Higgsing a linear combination, even though only four-fermion terms are present \cite{Ftnote5}. From energetics, it is expected that the rank is maximized $r=\min(K_\chi,K_\psi)$. The order parameter \pref{eqGH} can support topological defects, even in the flavor-symmetric case $K_\psi=K_\chi>1$ \cite{Huang2021}. The non-trivial homotopy group of $G/H$ implies the presence of stable defects, making the strong coupling phase a natural host for exotic disordered states, ranging from quantum paramagnets to phases with intrinsic topological order.
 
\emph{Summary} -- We have connected the concepts of conformal window and dynamic mass generation to the lineage of chirally stabilized fixed points and generalized the latter to arbitrary dimension and finite $N$. This is done by developing a generalization of Wilsonian OPE-RG technique to arbitrary $d$ and applying it to a two-species NJL-type model. Previous OPE-RG studies in $d>2$ focused on scalar theories \cite{Pagani2020}, whereas current–current interactions have been treated only with dimensional regularization. Our shell/OPE approach reproduces known $d=2$ results, extends them to $d>2$ and, crucially, keeps all power-divergent terms needed for mass deformations. Although we have focused on $d=2+\eps$,  the continuity indicates that the fixed points exist beyond these limits. We find that symmetry-breaking masses are not merely relevant in the RG sense; they dramatically alter the short-distance flow, pushing the system to strong coupling and triggering dynamical mass generation \cite{Semenoff1994,Ge2024}. 
Lastly, the rich manifold of the order parameter reinforces the notion that CFT + deformation is a route to topology.

This study was inspired by the dynamical large-N results \cite{Ge2024}. However, several differences need to be highlighted. In \cite{Ge2024} the conformal window at $d=3$ extended to $\kappa_\psi<4$, and the $\kappa_\chi$-threshold for symmetry-breaking dynamic mass generation was negligible. We note that the continuum limit of the microscopic model treated in \cite{Ge2024} requires both $\lambda_K{\cal K}_\psi^{A\mu\nu}{\cal K}^\chi_{A\mu\nu}$ and $\lambda_J{\cal J}_\psi^{A\mu}{\cal J}^\chi_{A\mu}$ interactions where ${\cal K}_\psi^{A\mu\nu}\equiv\bar\psi T^A\gamma^\mu \tau^\nu\psi$.  Such a problem can be treated using the same tools developed here.
In the massless limit, the flow topology resembles that of multi-channel Kondo fixed points. This suggests that the U(1) symmetry breaking (e.g., by superconducting proximity) could enable access to the strong coupling in over-screened impurity problems \cite{Komijaniqubit,Kattel2025}. A detailed investigation of these open questions is left for the future.

\emph{Acknowledgement} -- We thank Y.~Ge and K.~Wang for valuable discussions. L.\,C.\,R.\,W.\, is partially supported by the US. Department of Energy grant DE-SC1019775.

\subsection*{End Matter}
Here we summarize the coefficients of the $f_i(d)$ and $g_i(d)$ for various dimensions, leaving the detailed calculations to the supplementary materials \cite{SM}. Pulling out the $N$-dependence as $f_i(d,N)=(1-N^{-2})\tilde f_i(d)$, the one-loop results are
\be
\tilde f_2(d)=\frac{1}{\Omega_d}, \qquad g_2(d)=\frac{1}{\Omega_d}(3-\frac{2}{d}).
\ee
The two-loop results are
\bea
\tilde f_3(d)&=&\frac{6}{3^{d/2}}\frac{\Omega_{d-1}}{\Omega_d^4}{\cal I}_{CB(BB)}(d),\\
g_3(d)&=&\frac{16}{6^{d/2}}\frac{\Omega_{d-1}}{\Omega_d^3}(3-\frac{2}{d}){\cal I}_{C(BB)}(d),\\
g'_3(d)&=&\frac{16\times 6}{3^{d/2}}\frac{\Omega_{d-1}}{\Omega_d^3}(1-\frac{2}{d}){\cal I}_{CC}(d),
\eea
where ${\cal I}(d)$ coefficients involve 2$\times$d-dimensional integrals over ${\cal B}'$ in Fig.\,\ref{Fig1}(a), defined as $\theta(a<r_{12}<R_{\rm IR})\theta(a<r_{23}<R_{\rm IR})\theta(a<r_{13}<R_{\rm IR})$. To isolate the leading UV divergent contribution of these integrals, we i) go to center of mass coordinate $\vec R$ and two relative distances $\vec v$ and $\vec w$, ii) take the derivative with respect to the short-distance cut-off $a$, iii) re-scale the coordinates to explicitly pull out the $a$-dependence, and then iv) send $a\to 0$. Finally, by choosing a combination of spherical and cylindrical coordinates, the integrals (which is effectively the $\vec r_3$ integral in Fig.\,\ref{Fig1}a) are brought to two dimensions, over the region ${\cal B}$ defines as $\theta(1<s)\theta(1<t)$ where
\be
s=\sqrt{x^2+y^2}\sim r_{13}, \quad t=\sqrt{(x+1)^2+y^2}\sim r_{23}.
\ee
In terms of these the $\vec r_3$ integral involves the summations
\bea
{\cal I}_{CB(BB)}(d)&\equiv&\int_{\cal B}\frac{dxdy}{s^d}\big[1+\frac{1}{s^{d-2}}+\frac{1}{t^{3d-4}}\big],\\
{\cal I}_{C(BB)}(d)&\equiv&\int_{\cal B} dx dy \frac{\abs{y}^{d-2}}{s^d}\big[  \frac{ 2x }{s^{2(d-1)}}  - \, \frac{x+s^2}{s^{d-1} t^{3d-2}}  \big],\nonumber
\\
{\cal I}_{CC}(d)&\equiv& \int_{\cal B} dx dy\,\frac{ \abs{y}^{d-2}}{s^{2d}} \big[ (x^2-y^2)+\frac{t^2-2(y/s)^2}{2t^{2d}}  \big].\nonumber
\eea
These integrals are evaluated numerically. In $d=2$ (sufficient for the $\eps$-expansion), we can additionally use the Stokes' theorem to reduce them to contour integrals over $\partial{\cal B}$. Note that $g'_3\propto\eps\to 0$. We find that $g_3\ll g_2$ and $f_3\ll f_2$ due to the phase-space suppression $\Omega_{d-1}\Omega_d^{-3}$.

\bibliography{Anyons2.bib}

\begin{thebibliography}{46}
\bibitem [{\citenamefont {{Note that by flavor-symmetric ($K_\psi=K_\chi$) and
  flavor-asymmetric ($K_\psi\neq K_\chi$) we refer to the number of flavors.
  The interaction coupling $\lambda$ is a scalar and always symmetric.
  }}()}]{Ftnote2}%
  \BibitemOpen
  \bibfield  {author} {\bibinfo {author} {\bibnamefont {{Note that by
  flavor-symmetric ($K_\psi=K_\chi$) and flavor-asymmetric ($K_\psi\neq
  K_\chi$) we refer to the number of flavors. The interaction coupling
  $\lambda$ is a scalar and always symmetric. }}},\ }\href@noop {}
  {}\BibitemShut {NoStop}%
\bibitem [{\citenamefont {{To have a closed OPE algebra in $d=3$, we would need
  to introduce another set of operators ${\cal Q}^A=\bar\psi T^A\psi$, which
  for simplicity we avoid in this paper.}}()}]{Ftnote1}%
  \BibitemOpen
  \bibfield  {author} {\bibinfo {author} {\bibnamefont {{To have a closed OPE
  algebra in $d=3$, we would need to introduce another set of operators ${\cal
  Q}^A=\bar\psi T^A\psi$, which for simplicity we avoid in this paper.}}},\
  }\href@noop {} {}\BibitemShut {NoStop}%
\bibitem [{\citenamefont {{Note that the O(4) PSV symmetry of the order
  parameter is independent of the mass term \cite{SM}, but in a microscopic
  model it is typically broken by UV kinetic terms that are not taken into
  account.}}()}]{Ftnote3}%
  \BibitemOpen
  \bibfield  {author} {\bibinfo {author} {\bibnamefont {{Note that the O(4) PSV
  symmetry of the order parameter is independent of the mass term \cite{SM},
  but in a microscopic model it is typically broken by UV kinetic terms that
  are not taken into account.}}},\ }\href@noop {} {}\BibitemShut {NoStop}%
\end{thebibliography}

\newpage
\setcounter{equation}{0}
\bw

\section*{\Large\bf Supplementary Materials}

This supplementary material provides proof of various statements made in the paper including an explicit computation of the beta function coefficients.
\tableofcontents

\section{\large\sc Non-interacting fermions in 2+1D}
A normal Dirac fermions in Graphene has two sublattices and two valleys. The Hamiltonian can be written as
\be
H=\sum_k\tilde\psi\dg\tilde{\cal H}\tilde\psi,\qquad \tilde\psi=\mat{\psi_{RA}\\\psi_{RB}\\\psi_{LA}\\\psi_{LB}} \qquad \tilde{\cal H}=\mat{v\vec k\cdot\sigma & \\ & -v\vec k\cdot\sigma^*}=vk_x\sigma^x\tau^z+vk_y\sigma^y, 
\ee
A useful modification is to do a rotation $i\sigma^y$ on left cone:
\be
H=\sum_k\psi\dg{\cal H}\psi,\qquad \psi=O\tilde\psi=\mat{\psi_{RA}\\\psi_{RB}\\\psi_{LB}\\-\psi_{LA}} \quad {\cal H}=\mat{v\vec k\cdot\sigma & \\ & v\vec k\cdot\sigma}=vk_x\sigma^x+vk_y\sigma^y, \quad -G^{-1}(z)=-z\bb 1+{\cal H}
\ee
The spectrum is two copies of $E=\pm v\sqrt{k_x^2+k_y^2}=\pm v\abs{k}$. The Euclidean Lagrangian is 
\be
{\cal L}=\psi\dg[-G^{-1}(i\omega_n)]\psi=\psi\dg[-i\omega_n\bb 1+vk_x\Gamma^1+vk_y\Gamma^2]\psi, \qquad \Gamma^1=\sigma^x, \quad \Gamma^2=\sigma^y
\ee
Lastly, we define $\bar\psi=\psi\dg\gamma^0$, and define $k_0=-\omega_n$. Then,
\be
\bar\psi\equiv \psi\dg\sigma^z \so {\cal L}=i\bar\psi[k_0\gamma^0+k_x\gamma^1+k_y\gamma^2]\psi, \qquad \gamma^0=\sigma^z, \quad \gamma^1=-i\sigma^z\sigma^x=\sigma^y, \quad \gamma^2=-i\sigma^z\sigma^y=-\sigma^x
\ee
where we have also set the velocity to 1 by re-scaling $x$ and $y$ axis. Note that we have $\gamma^\mu=\gamma^\mu\otimes\tau^0$.\\
To go to the real-space we use 
\be
\psi(k)=\int{d^dr}e^{-i\vec k\cdot\vec r}\psi(r), \andd \psi(r)=\int{\frac{d^dk}{(2\pi)^d}}e^{i\vec k\cdot\vec r}\psi(k)
\ee
to conclude that ${\cal L}=\bar\psi (i\gamma^\mu k_\mu)\psi$ or ${\cal L}=\bar\psi (\gamma^\mu\partial_\mu)\psi=\bar\psi \cancel\partial\psi$

\section{\large\sc Green's function}
\subsection{The massless case}
Here we prove that the correct Green's function is
\be
-G^{-1}=\cancel\partial, \qquad G(\vec r)\equiv\braket{-{\cal R}\psi(\vec r)\bar\psi}=-\frac{\cancel r}{r^d\Omega_d}\equiv -\cancel rG(r) \so (-\cancel\partial)(-\frac{\cancel r}{r^d\Omega_d})=\delta^d(\vec r).\label{eq21}
\ee
We use the usual (hereafter implicit) \emph{radial ordering} ${\cal R}$ to define $G(\vec r)$. We have also defined $G(r)\equiv 1/r^d\Omega_d$. To show that the Green's function is correct, we write
\be
\partial_\mu(\frac{r_\nu}{r^d})=\frac{1}{r^d}(\delta_{\mu\nu}-\frac{r_\mu r_\nu}{r^2}d) \so \gamma^\mu\gamma^\nu \partial_\mu(\frac{r_\nu}{r^d})={\delta^{\mu\nu}}(\delta_{\mu\nu}-\frac{r_\mu r_\nu}{r^2}d)=(1-\frac{r^2}{r^2})d
\ee
where we used that RHS of the first equation is symmetric and only picks the symmetric part of $\gamma^\mu\gamma^\nu=\delta^{\mu\nu}\bb 1+\sigma^{\mu\nu}$. Therefore this is zero at any $r\neq0$. To show that it is a delta function, we integrate
\be
\frac{1}{\Omega_d}\gamma^\mu\gamma^\nu\int_B{d^dr}\partial_\mu \frac{r_\nu}{r^d}=\frac{1}{\Omega_d}\gamma^\mu\gamma^\nu\oint_{\partial B}{d\Omega_d}r^{d-1}\hat r_\mu \frac{r_\nu}{r^d}=\frac{1}{\Omega_d}\gamma^\mu\gamma^\nu\oint_{\partial B}{d\Omega_d}\hat r_\mu \hat r_\nu=1
\ee
where we used Stokes theorem. Therefore, we conclude that \pref{eq21} is correct.
\subsection{The massive case}
The Green's function for fermions $G(\vec r)\equiv\braket{-\psi(\vec r)\bar\psi}$ is related to that of a corresponding boson
\be
-G_f^{-1}(\vec r)=\cancel\partial+m \so G_f(\vec r)=-(\cancel\partial+m)^{-1}=(\cancel\partial-m)G_b(r), \qquad G_b(r)\equiv(-\cancel\partial^2+m^2)^{-1}
\ee
where $G_b$ is the Green's function of massive scalar theory.  The latter can be found in momentum space
\be
G_b(\vec r)=\int{\frac{d^dk}{(2\pi)^d}}e^{i\vec k\cdot\vec r} \tilde G_b(k), \qqquad \tilde G_b(k)=\int{d^dr}G_b(r)e^{-i\vec k\cdot\vec r}=
\frac{1}{m^2+k^2}
\ee
from which we find
\be
G_b(\vec r)=-\frac{(2\pi)^{d/2}}{(2\pi)^d}\int\frac{k^{d-1}dk}{k^2+m^2}\frac{J_{d/2-1}(kr)}{(kr)^{d/2-1}}=-\frac{1}{(2\pi)^{d/2}}\Big(\frac{m}{r}\Big)^{d/2-1}K_{d/2-1}(mr)
\ee
Note that $K_\nu(x)$ with $\nu=d/2-1$ has the asymptotics
\be
\nu\neq 0:\qqquad \lim_{x\to 0}K_\nu(x)\sim \frac{1}{2}\Big\{\Gamma(\nu)\Big(\frac{x}{2}\Big)^{-\nu}+\Gamma(-\nu)\Big(\frac{x}{2}\Big)^\nu\Big\},\qquad \lim_{x\to\infty}K_\nu(x)=\sqrt{\frac{\pi}{2x}}e^{-x}
\ee
For $\nu>0$, the second term is subleading and can be dropped, but for $\nu\to 0$ (relevant to $d=2$) we can use this along with $\Gamma(\pm\eps\to 0)=\pm 1/\eps-\gamma_E+\pi^2\eps^2/12 +\dots$, or more directly,
\be
\nu=0:\qqquad \lim_{x\to 0}K_0(x)=-[\log(x/2)+\gamma_E],\qquad \lim_{x\to\infty}K_0(x)=\sqrt{\frac{\pi}{2x}}e^{-x}
\ee
The fermionic Green's function is
\be
G_f(\vec r)=(\gamma^\mu\partial_\mu-m)G_b(\vec r)=( \frac{\cancel{r}}{r}\frac{d}{dr} -m)G_b(\vec r)
\ee
Using $K'_\nu(x)=-K_{\nu+1}(x)+\frac{\nu}{x}K_\nu(x)$ and $K'_\nu(x)=-K_{\nu-1}(x)-\frac{\nu}{x}K_\nu(x)$
we find
\be
G_f(\vec r)=-G(r)\Big[{\cancel r}+rn(mr)\Big],\qquad G(r)\equiv\frac{g(mr)}{\Omega_d r^d} \qquad n(mr)\equiv\frac{K_\nu(mr)}{K_{\nu+1}(mr)}, \qquad g(x)\equiv \frac{x^{\nu+1}K_{\nu+1}(x)}{2^\nu\Gamma(\nu+1)} 
\ee
In the limit of $x=mr\ll 1$ the second term inside the square bracket is tricky:
\be
\lim_{x\to 0}n(x)=\frac{\Gamma(\nu)(x/2)^{-\nu}+\Gamma(-\nu)(x/2)^\nu}{\Gamma(\nu+1)(x/2)^{-\nu-1}+\Gamma(-\nu-1)(x/2)^{\nu+1}}=x\left\{\matl{1/2\nu & \nu>0 \\ -[\gamma_E+\log (x/2)] \qquad & \nu=0}\right.
\ee
In the limit of $x=mr\ll 1$ we find 
\be
G_f(\vec r)=-\frac{1}{\Omega_d}\frac{1}{r^{d}}\Big[{\cancel r}+r n(mr)\Big], \qquad n(mr)=mr\left\{\matn{1/(d-2) & d>2 \\ -\gamma_E-\log(mr/2)\qquad & d=2}\right.
\ee
On the other hand in the limit of $x=mr\gg 1$, using
\be
\lim_{x\to\infty}K_\nu(x)\sim \sqrt{\frac{\pi}{2x}}e^{-x}\Big[1+\frac{4\nu^2-1}{8x}+O(x^{-2})\Big] \so \frac{K_\nu(x)}{K_{\nu+1}(x)}=1-\frac{d-1}{2x}
\ee
we find
\be
G_f(\vec r)=-\frac{1}{2}\frac{\cancel r+r}{(2\pi r/m)^{(d-1)/2}}e^{-mr}
\ee

\section{\large\sc Operator Product Expansion}
Here, we derive OPE relations for non-interacting Dirac fermions in $d=2\dots 4$ in both massless and massive cases.

Quite generally, can write $G(\vec r)=\braket{-\psi(\vec r)\bar\psi}=-G(r)r_\alpha\gamma^\alpha$ where $G(r)$ only depends on the magnitude of $r$. However, for $r_\alpha$ we allow for even coordinates $(-r)_\alpha\neq r_\alpha$ to be able to incorporate the mass term. The only possible contractions are
\bea
(\bar\psi \gamma^\mu T^A\psi)_{\vec r}(\bar\psi \gamma^\nu T^B\psi)_{0}
&\sim& 
G^2(r)(-r_\alpha)(-r)_\beta\tr{\gamma^\mu\gamma^\alpha\gamma^\nu\gamma^\beta}\delta^{AB}+G(r)\bar\psi\Big(r_\alpha \gamma^\mu \gamma^\alpha\gamma^\nu T^AT^B+(-r)_\alpha\gamma^\nu \gamma^\alpha\gamma^\mu T^BT^A\Big)\psi\nonumber\\
&&\hspace{4cm}+G(r)r^\beta\bar\psi \Big(r_\alpha \gamma^\mu \gamma^\alpha\gamma^\nu T^AT^B-(-r)_\alpha\gamma^\nu \gamma^\alpha\gamma^\mu T^BT^A\Big)\partial_\beta\psi
\eea
We use $T^AT^B=\frac{1}{2}(if^{ABC}T^C+{\cal D}^{AB})$. We also have the relations
\bea
&&d=2:\quad \gamma^0=\sigma^x,\quad\gamma^1=-\sigma^y, \quad \gamma^5=\sigma^z,\quad\gamma^\mu\gamma^\nu=\delta^{\mu\nu}\bb 1-i\eps^{\mu\nu}\gamma^5, \\
&&d=3:\quad \gamma^0=\sigma^z,\quad\gamma^1=\sigma^y, \quad \gamma^2=-\sigma^x,\quad\gamma^\mu\gamma^\nu=\delta^{\mu\nu}\bb 1+i\eps^{\mu\nu\lambda}\gamma^\lambda, \quad
\gamma^\mu\gamma^\nu\gamma^\rho=\delta^{\mu\nu}\gamma^\rho+\delta^{\nu\rho}\gamma^\mu-\delta^{\mu\rho}\gamma^\nu+i\eps^{\mu\nu\rho}\bb 1\nonumber
\eea
and both satisfying (as long as the usual gamma matrices are involved)
\be
\tr{\gamma^\mu\gamma^\alpha\gamma^\nu\gamma^\lambda}=4[\delta^{\mu\alpha}\delta^{\lambda\nu}+\delta^{\mu\lambda}\delta^{\alpha\nu}-\delta^{\alpha\lambda}\delta^{\mu\nu}]
\ee
Note that $\gamma^\mu$, $\gamma^\nu$ come from currents and $\gamma^\alpha$ from the Green's function.
We can use the fact that
\be
r_\alpha\gamma^\mu\gamma^\alpha\gamma^\nu=R^{\mu\nu}\bb 1+B^{\mu\nu}_\lambda\gamma^\lambda, \quad R^{\mu\nu}=\frac{r_\alpha}{4}\tr{\gamma^\mu\gamma^\alpha\gamma^\nu}, \quad B^{\mu\nu}_\lambda=\frac{r_\alpha}{4}\tr{(\gamma^\mu\gamma^\alpha\gamma^\nu)\gamma_\lambda}
\ee
To be able to define this we include $-\gamma^5$ in $\gamma^\lambda$ for the $d=2$ case.
In terms of
\bea
C^{\mu\nu}&=&\frac{1}{4}(-r_\alpha)(-r)_\beta\tr{\gamma^\mu\gamma^\alpha\gamma^\nu\gamma^\beta}=(-r_\alpha)(-r)_\beta[\delta^{\mu\alpha}\delta^{\beta\nu}+\delta^{\mu\beta}\delta^{\alpha\nu}-\delta^{\alpha\beta}\delta^{\mu\nu}]\\ 
B^{\mu\nu}_{\pm\lambda}&=&\frac{1}{8}\tr{(r_\alpha \gamma^\mu \gamma^\alpha\gamma^\nu \mp(-r)_\alpha\gamma^\nu \gamma^\alpha\gamma^\mu)\gamma^\lambda}=\frac{r_\alpha\mp(-r)_\alpha}{2}[\delta^{\mu\alpha}\delta^{\lambda\nu}+\delta^{\mu\lambda}\delta^{\alpha\nu}-\delta^{\alpha\lambda}\delta^{\mu\nu}]\\
R_\pm^{\mu\nu}&=&\frac{1}{8}\tr{r_\alpha \gamma^\mu \gamma^\alpha\gamma^\nu \mp (-r)_\alpha\gamma^\nu \gamma^\alpha\gamma^\mu}
\eea
In the massless case, we have [note that $\gamma^\lambda=\gamma^5$ contribution would not contribute in the massless case]. In the massive case, $\alpha$, $\beta$ and $\lambda$ can refer to $\gamma^m=\bb 1$. Also $\lambda$ could refer to $\gamma^5$ for $d=2$.
We can write
\bea
{\cal J}^{A\mu}(\vec r){\cal J}^{B\nu}(0)&\sim& 4G^2(r)C^{\mu\nu}(\vec r)\delta^{AB}+G(r)\Big[if^{ABC}\bar\psi T^C\Big(B^{\mu\nu}_{+\lambda}\gamma^\lambda+R^{\mu\nu}_+\Big)\psi+\bar\psi{\cal D}^{AB}\Big(B^{\mu\nu}_{-\lambda}\gamma^\lambda+R^{\mu\nu}_-\Big)\psi\\
&&\hspace{4cm}-if^{ABC}r^\beta\bar\psi T^C\Big(B^{\mu\nu}_{-\lambda}\gamma^\lambda+R^{\mu\nu}_-\Big)\partial_\beta\psi-r^\beta\bar\psi{\cal D}^{AB}\Big(B^{\mu\nu}_{+\lambda}\gamma^\lambda+R^{\mu\nu}_+\Big)\partial_\beta\psi\nonumber
\eea
Note that the derivative terms appear with a minus sign. In the massless case, we have
\be
C^{\mu\nu}=2r_\mu r_\nu-\delta^{\mu\nu}r^2, \qquad B^{\mu\nu}_{+\lambda}=
r^\mu \delta^\nu_\lambda+r^\nu\delta^\mu_\lambda-r_\lambda \delta^{\mu\nu},
\ee
Additionally, for $d=2$ and $d=4$, we have $R_\pm^{\mu\nu}=0$, whereas in $d=3$ still $R_+^{\mu\nu}=0$, holds but $R_-^{\mu\nu}=ir_\alpha \eps^{\mu\alpha\nu}$. Therefore, we find the OPE mentioned in the paper
\bea
{\cal J}^{A\mu}(\vec r){\cal J}^{B\nu}(0)&\sim& [2G(r)]^2C^{\mu\nu}(\vec r)\delta^{AB}+G(r)B^{\mu\nu}_{+\lambda}\Big[if^{ABC}{\cal J}^{C\lambda}-r^\beta\bar\psi{\cal D}^{AB}\gamma^\lambda\partial_\beta\psi\Big]\nonumber
\eea
In the massive case $m\neq 0$, we have $B^{\mu\nu}_{+\lambda}$ remain the same and still $B^{\mu\nu}_{-\lambda}=0$ but we get the following corrections
\be
\tilde C^{\mu\nu}(\vec r)=2r^{\mu}r^\nu-r^2(1+n^2)\delta^{\mu\nu}, \qquad R^{\mu\nu}_-=nr\delta^{\mu\nu}, \qquad R^{\mu\nu}_+=0.
\ee
This leads to the OPE
\bea
{\cal J}^{A\mu}(\vec r){\cal J}^{B\nu}(0)&\sim& 4G^2(r)\tilde C^{\mu\nu}(\vec r)\delta^{AB}-G(r)\Big\{B^{\mu\nu}_{+\lambda}(\vec r)\Big[if^{ABC} {\cal J}^{C\lambda}-r^\beta \bar\psi{\cal D}^{AB}\gamma^\lambda\partial_\beta\psi\Big]\\
&&\hspace{4cm}+rn(mr)\delta^{\mu\nu}\Big[\bar\psi{\cal D}^{AB}\psi-if^{ABC}rn(mr)r^\beta\bar\psi T^C\partial_\beta\psi\Big]\Big\}\nonumber
\eea
The modifications include $C\to\tilde C$ and the entire second line. In the second line, the first term is responsible for mass renormalization. The second term always vanishes in two-loop calculations because it is accompanied with a $\delta^{AB}$ from the other fermion sector.
\subsection{Two-dimensional OPEs in complex coordinates}
We transform 
\be
C^{\mu\nu}(\vec r)=2r^\mu r^\nu -r^2\delta^{\mu\nu}, \qqquad B^{\mu\nu}_\lambda(\vec r)=r^\mu \delta^\nu_\lambda+r^\nu\delta^\mu_\lambda-r_\lambda\delta^{\mu\nu}
\ee
in $d=2$ to the light-cone coordinates
\be
\matn{z^1=z=r^1+ir^2 \\ z^2=\bar z=r^1-ir^2} \so {J^\mu}_{\mu'}=\frac{\partial z^\mu}{\partial r^{\mu'}}=\mat{1 & i \\ 1 & -i}, \qquad {(J^{-1})^{\mu'}}_\mu=\frac{\partial r^{\mu'}}{\partial z^{\mu}}=\frac{1}{2}\mat{1 & 1 \\ -i & i}
\ee
We can also compute the metric
\be
\matn{\partial_{z^1}=\frac{1}{2}(\partial_{r^1}-i\partial_{r^2}) \\ \partial_{z^1}=\frac{1}{2}(\partial_{r^1}+i\partial_{r^2})}, \quad
\matn{dz^1=dr^1+idr^2 \\ dz^2=dr^1-idr^2},\quad ds^2=dr_1^2+dr^2=dz^1dz^2, \quad g_{z^iz^j}=\frac{1}{2}\mat{0 & 1 \\ 1 & 0}, \quad g^{z^iz^j}=\mat{0 & 2 \\ 2 & 0}.
\ee
The calculation to be done for $B$ is
\be
\tilde B^{\mu\nu}_\lambda(z^1,z^2)=(\frac{\partial z^\mu}{\partial r^{\mu'}})(\frac{\partial z^\nu}{\partial r^{\nu'}})(\frac{\partial r^{\lambda'}}{\partial z^{\lambda}})B^{\mu'\nu'}_{\lambda'}(r^1,r^2)={J^{\mu}}_{\mu'}{J^{\nu}}_{\nu'}{(J^{-1})^{\lambda'}}_{\lambda}B^{\mu'\nu'}_{\lambda'}(r^1,r^2)
\ee
For a fixed $\mu,\mu'$ part of this calculation can be done in the matrix form:
\bea
&&{B^{1,\nu}}_\lambda=r^1\mat{1 \\ & 1}+\mat{r^1 & 0 \\ r^2 & 0}-\mat{r_1 & r_2 \\ 0  &0}=\mat{r^1 & -r^2 \\ r^2 & r^1}
\\
&&
{B^{2,\nu}}_\lambda=r^2\mat{1 \\ & 1}+\mat{0 & r^1 \\ 0 & r^2}-\mat{0  &0 \\ r_1 & r_2}=\mat{r^2 & r^1 \\ -r^1 & r^2}
\eea
Then, we have
\be
{(B^{\mu'=1})^{\nu}}_\lambda=J\times B^{\mu'=1}\times J^{-1}=\mat{1 & i \\ 1 & -i}\mat{r^1 & -r^2 \\ r^2 & r^1}\frac{1}{2}\mat{1 & 1 \\ -i & i}=\mat{z^1 \\ & z^2}
\ee
\be
{(B^{\mu'=2})^{\nu}}_\lambda=J\times B^{\mu'=2}\times J^{-1}=\mat{1 & i \\ 1 & -i}\mat{r^2 & r^1 \\ -r^1 & r^2}\frac{1}{2}\mat{1 & 1 \\ -i & i}=\mat{-iz^1 \\ & iz^2}
\ee
Out of these, we finally have
\be
{(\tilde B^{\mu=1})^{\nu}}_\lambda=(J^1)_1{(B^{\mu'=1})^{\nu}}_\lambda+(J^1)_2{(B^{\mu'=2})^{\nu}}_\lambda=(B^{\mu'=1})+i(B^{\mu'=2})=2\mat{z^1 \\ & 0}
\ee
\be
{(\tilde B^{\mu=1})^{\nu}}_\lambda=(J^2)_1{(B^{\mu'=1})^{\nu}}_\lambda+(J^2)_2{(B^{\mu'=2})^{\nu}}_\lambda=(B^{\mu'=1})-i(B^{\mu'=2})=2\mat{0 \\ & z^2}
\ee
So, we conclude that the tensor $\tilde B^{\mu\nu}_\lambda(z^1,z^2)$ has only two non-zero elements: $\tilde B^{11}_1=2z^1$ and $\tilde B^{22}_2=2z^2$.
\sepline
Next, we look at $C^{\mu\nu}$. 
\be
C^{\mu\nu}=\mat{(r^1)^2-(r^2)^2 & 2r^1r^2\\ 2r^1r^2 & (r^2)^2-(r^1)^2}, \qqquad J=\mat{1 & i \\ 1 & -i}, \qquad
J^{-1}=\frac{1}{2}\mat{1 & 1 \\ -i & i}
\ee
We have
\be
\tilde C^{\mu\nu}=(J^\mu)_{\mu'} (J^\nu)_{\nu'} C^{\mu'\nu'} \so \tilde C=\mat{C^{11}+i(C^{12}+C^{21})+i^2C^{22}  & C^{11}+iC^{12}-iC^{21}+i(-i)C^{22} \\ C^{11}-iC^{12}+iC^{21}+(-i)iC^{22} & C^{11}-i(C^{12}+C^{21})+(-i)^2C^{22}}
\ee
giving
\be
\tilde C^{\mu\nu}=\mat{(r^1+ir^2)^2 \\ & (r^1-ir^2)^2 }=\mat{(z^1)^2 \\ & (z^2)^2}\quad\to\quad \tilde C_{\mu\nu}=g_{\mu\mu'}g_{\nu\nu'}\tilde C^{\mu'\nu'}=\frac{1}{4}\mat{\tilde C^{22} & \tilde C^{21} \\ \tilde C^{12} & \tilde C^{11}}=\frac{1}{4}\mat{(z^2)^2 \\ & (z^1)^2}
\ee
Note that both $\tilde B$ and $\tilde C$ match our expectations.

\section{\large\sc Beta function}
\subsection{One-loop RG}
The Euclidean action in $d$ spacetime dimensions is $S=\int{d^dr}{\cal L}$ where:
\begin{equation}
{\cal L}=\bar\psi_{\alpha a} \cancel\partial \, \psi_{\alpha a}+\bar\chi_{\alpha b}\cancel\partial\chi_{\alpha b}+\lambda \, {\cal J}^{A,\mu}_\psi{\cal J}_{A,\mu}^\chi
\end{equation}
The RG procedure is as follows. We divide $r$'s of each current into the fast ($a<r<a+da$) and slow ($r>a+da$) modes, where $a$ is a short distance cutoff. We do the perturbative expansion in $\lambda$, integrate out the fast modes and then we re-scale the position $r \to (1+d\ell)r$, with $d \ell = da/a$. This gives us the renormalization of the coupling constant as:
\begin{equation}
    \lambda^\prime = \lambda \, (1+d\ell)^d \frac{Z_\lambda}{Z_\psi \, Z_{\chi}}(1+d \ell)^{1-d} (1+d \ell)^{1-d}  = \lambda (1+ d \ell)^{2-d}(1+d\ell)^{y_\lambda - y_\psi -y_\chi} = \lambda + \lambda (-\epsilon +y_\lambda - y_\psi -y_\chi ) d \ell
\end{equation}
Where $\lambda^\prime$ is the new, re-scaled coupling constant, $Z_{\lambda} =(1+y_\lambda)d\ell$ is the renormalization of the interaction, and $Z_{\psi},Z_{\chi}$ are wave function renormalizations for each of the fields. The way we perform the fast mode integration is by utilizing OPE's of the currents to extract all possible contractions and integrate out the fast modes in a range from $a$ to $a+da$ This step is a subject of the following few sections.
OPE's of the currents (in the massless case) are given by:
\begin{equation}
\begin{split}
    {\cal J}^{A,\mu}_{\psi} (\vec{r}) {\cal J}^{B ,\nu}_{\psi} (\vec{0}) = k_{\psi} \ \delta^{AB} \frac{(2g_d)^2}{|r|^{2d}}  C^{\mu \nu}(\vec{r}) + i f^{ABC}\frac{g_d}{|r|^d} B^{\mu \nu}_{\lambda} (\vec{r}) \ {\cal J}_{\psi}^{C,\lambda}- \frac{g_d}{|r|^d} r_\beta B^{\mu \nu}_{\lambda} (\vec{r}) \bar{\psi}  \, {\cal D^{AB}} \gamma^\lambda \partial_\beta \psi \\
    {\cal J}^{A,\mu}_{\chi} (\vec{r}) {\cal J}^{B ,\nu}_{\chi} (\vec{0}) = k_{\chi}\delta^{AB} \frac{(2g_d)^2}{|r|^{2d}}  C^{\mu \nu}(\vec{r}) + i f^{ABC}\frac{g_d}{|r|^d} B^{\mu \nu}_{\lambda} (\vec{r}) \ {\cal J}_{\chi}^{C,\lambda}- \frac{g_d}{|r|^d} r_\beta B^{\mu \nu}_{\lambda} (\vec{r}) \bar{\chi}  \, {\cal D^{AB}} \gamma^\lambda \partial_\beta \chi
\end{split}
\end{equation}
Where we have explicitly used $G(r) \equiv \frac{1}{\Omega_d|r|^d}$ and  $g_d=1/\Omega_d$ and :
\begin{equation}
    \begin{split}
        C^{\mu \nu}(\vec{r}) = 2 r^{\mu} r^{\nu} - \delta^{\mu \nu}r^2, \qquad B^{\mu \nu}_{\lambda} (\vec{r}) = r^{\nu} \delta^{\mu}_{\lambda} + r^{\mu} \delta^{\nu}_{\lambda} - r_{\lambda} \delta^{\mu \nu}
    \end{split}
\end{equation}
Here we omit the $+$ label in the $B_{+ \lambda}^{\mu \nu}$ to not burden the notation.
\newline
\subsubsection*{Contribution to $y_\lambda$:}
According to the above OPE to one loop we only have the following contribution to the interaction renormalization (to $y_\lambda$):
\begin{equation}
\begin{split}
    &-f^{ABC} f^{ABD} \frac{ B^{\mu \nu}_{\rho}(\vec{r}_{12})B_{\mu \nu}^{\xi}(\vec{r}_{12})}{(r_{12})^{2d}}{\cal J}^{C,\rho}_{\psi} (\vec{r}_1) {\cal J}_{\chi}^{D,\xi}(\vec{r}_{1}) = -C_{2} \frac{ B^{\mu \nu}_{\rho}(\vec{r}_{12})B_{\mu \nu}^{\xi}(\vec{r}_{12})}{(r_{12})^{2d}}{\cal J}^{C,\rho}_{\psi} (\vec{r}_1) {\cal J}_{\chi}^{C,\xi}(\vec{r}_{1})
\end{split}
\end{equation}
Where we have used $f^{ABC} f^{ABD}= C_2 \delta^{CD}$ and $C_2$ is a quadratic Casimir of the group. We can simplify it further, by using the definition and properties of $B$ tensors:
\begin{equation}
    B^{\mu \nu}_{\rho}(\vec{x})B_{\mu \nu}^{\xi}(\vec{x}) =  x^2 (3-\frac{2}{d}) \delta^{\xi}_{\rho}
\end{equation}
Therefore, the integral becomes (after going to center of mass coordinates):
\begin{equation}
    -\lambda^2 (g_{d})^2 C_2  \, \bigg (3-\frac{2}{d} \bigg ) \, I_2 (a) \int d^d \vec{R}{\cal J}^{C,\rho}_{\psi} (\vec{R}) \,  {\cal J}_{\chi}^{C,\rho}(\vec{R})
\end{equation}
With:
\begin{equation}
    I_2 (a) = \int d^d \vec{r} \ \frac{1}{r^{2d-2}} \Theta (a<r<R), \qquad
    \partial_a I_2 = \int d^d \vec{r} \ \frac{\delta(r-a)}{r^{2d-2}} =  \frac{\Omega_d}{a^{d-1}}, \qquad \to \qquad I_2 = \Omega_d a^{2-d} d\ell    
\end{equation}
Where we have defined $d\ell = da/a$. There is a minus sign coming from bringing the term in the exponential. Therefore, we have that the $y_\lambda$ to one-loop is:
\begin{equation}
    y_\lambda = \bigg (3-\frac{2}{d} \bigg )\frac{C_2}{\Omega_d} \, \tilde{\lambda}^{}  \qquad Z_{\lambda}=1+y_\lambda \, d \ell\ >1 
\end{equation}
Where we have defined $\tilde{\lambda} = \lambda/a^{2-d}$. And $Z_{\lambda}$ is renormalization of the interaction ($\lambda^\prime = Z_{\lambda} \times \lambda$), with $\lambda^\prime$ being the new re-scaled coupling constant.
\newline
\subsubsection{Contribution to $y_\psi$}
According to the OPE the term that contributes to the wave function renormalization of $\psi$ is given by:
\begin{equation}
\begin{split}
    &-k_{\chi} \delta^{AB} \frac{ r_b B^{\mu \nu}_{\rho}(\vec{r}_{12})C_{\mu \nu}(\vec{r}_{12})}{(r_{12})^{3d}}\bar{\psi}  \, {\cal D}^{AB} \gamma^\rho \partial_b \psi= -k_{\chi} \frac{(N_f^2-1)}{N_f}  \frac{ r_b B^{\mu \nu}_{\rho}(\vec{r}_{12})C_{\mu \nu}(\vec{r}_{12})}{(r_{12})^{3d}}\bar{\psi}  \, \gamma^\rho \partial_b \psi
\end{split}
\end{equation}
Where we have used that ${\cal D}^{AB} = (1/2) \{ T^A, T^B \}$ where $T$'s are group generators. For $SU(N)$ we have   ${\cal D}^{AA} = \sum^{N^2-1}_{A} 1/N$ which gives ${\cal D}^{AA} = (N^2-1)/N$ which we use in the following.
\begin{equation}
    B^{\mu \nu}_{\rho}(\vec{x})C_{\mu \nu}(\vec{x}) = d\, x^2 x_\rho, \qquad x_b \, B^{\mu \nu}_{\rho}(\vec{x})C_{\mu \nu}(\vec{x}) = d\, x^2 x^b x_\rho \to x^4 \delta^{b}_{\rho}
\end{equation}
Therefore we have (again after going to the center-of-mass coordinates):
\begin{equation}
    -\lambda^2 (2g_d)^2 k_{\chi}\frac{(N_f^2-1)}{N_f} I_{\psi} \int d^d \vec{R} \bar{\psi} \gamma^{\rho}\partial_{\rho} \psi, \qquad
        I_{\psi} = \int d^d \vec{r} \frac{1}{r^{3d-2}} \theta(a<r<R)
\end{equation}
Taking the derivative
\begin{equation}
    \partial_a I_{\psi} = \int d^d \vec{r} \frac{\delta(r-a)}{r^{3d-4}} = \frac{\Omega_d}{a^{2d-3}} \qquad \to \qquad  I_{\psi} = \Omega_d a^{4-2d} d\ell
\end{equation}
Therefore, (after bringing it back into the exponential, and incuring an extra minus sign) we have:
\begin{equation}
    y_{\psi} = \frac{k_{\chi}}{\Omega_d} \frac{(N^2-1)}{N} \tilde{\lambda}^2(a), \qquad Z_{\psi}  = 1+y_\psi \, d\ell >1
\end{equation}
We get the same result for $y_\chi$ if we do the replacement $k_{\psi} \to k_{\chi}$. Inspecting this contribution we can see that it is of order $\tilde{\lambda}^2$ which means that in the expression for the beta function:
\begin{equation}
    \beta \equiv \frac{d\tilde{\lambda}}{d \ell} = -\epsilon \tilde{\lambda} + \tilde{\lambda} (y_\lambda-y_\psi - y_\chi) 
\end{equation}
it contributes to the two loop order (order $\tilde{\lambda}^3$). Therefore to one-loop we only have $Z_{\lambda}$. Lets look at more carefuly at the beta functiona to this order just to set up the convention for obtaining it: 
\begin{equation}
    \lambda^\prime =  \lambda + \lambda (-\epsilon +y_\lambda ) d \ell, \qquad \beta \equiv \frac{d\tilde{\lambda}}{d\ell} = -\epsilon \tilde{\lambda} + g_2 \tilde{\lambda}^2, \qquad g_2=(3-2/d)\frac{C_2}{\Omega_d}
\end{equation}
In the action in fact our coupling constantis rescaled by $N$. If we do the change $\tilde{\lambda} \to \tilde{\lambda}/N$ in the above exoression and using the fact that $C_2=N$ we arrive at the beta function:
\begin{equation}
    \beta  = -\epsilon \tilde{\lambda} + g_2 \tilde{\lambda}^2, \qquad g_2=(3-2/d)\frac{1}{\Omega_d}
\end{equation}

\subsection{Two-Loop RG}
 We can now look at the two-loop order ($\sim \lambda^3$). To calculate the $\beta$-function to this order we only need the two-loop contribution to $Z_\lambda$. The two-loop contribution to the beta function coming from the wave function renormalization has been previously calculated ($Z_\psi$ and $Z_{\chi}$ of the previous section). In order to count possible contractions of three currents (which is needed at this order) we can systematically write all of them in the form of two matrices
\begin{equation}
\begin{split}
\hat{\cal \chi} \equiv {\cal J}^{A,\mu}_{\chi} (\vec{r}_1){\cal J}^{B,\nu}_{\chi} (\vec{r}_2){\cal J}^{C,\rho}_{\chi} (\vec{r}_3) \qquad \qquad \qquad \qquad  \qquad \qquad \qquad \qquad  \\
\\
{\small \begin{bmatrix}
    &k_\chi \frac{\delta^{AB}C^{\mu \nu}(\vec{r}_{12})}{(r_{12})^{2d}}{\cal J}^{C,\rho}_{\chi} (\vec{r}_3)&k_\chi \frac{\delta^{AC}C^{\mu \rho}(\vec{r}_{13})}{(r_{13})^{2d}}{\cal J}^{B,\nu}_{\chi} (\vec{r}_2)& k_\chi \frac{\delta^{BC}C^{\nu \rho}(\vec{r}_{23})}{(r_{23})^{2d}}{\cal J}^{A,\mu}_{\chi} (\vec{r}_1) \\
    \\
   & -  \frac{f^{ABD} f^{DCE} B^{\mu \nu}_{\lambda}(\vec{r}_{12})B^{\lambda \rho}_{\xi}(\vec{r}_{23}) }{(r_{12}\ r_{23})^d} {\cal J}_{\chi}^{E,\xi}(\vec{r}_{3}) & - \frac{f^{ACD}f^{DBE}B^{\mu \rho}_{\lambda}(\vec{r}_{13})B^{\lambda \nu}_{\xi}(\vec{r}_{32})}{(r_{13} \ r_{32})^d} {\cal J}_{\chi}^{E,\xi}(\vec{r}_{3})& - \frac{f^{BCD}f^{DAE} B^{\nu \rho}_{\lambda}(\vec{r}_{23})B^{\lambda \mu}_{\xi}(\vec{r}_{31}) }{(r_{23} \ r_{31})^d}{\cal J}_{\chi}^{E,\xi}(\vec{r}_{3})\\
   \\
   &  -i\frac{f^{ABD}(r_{23})_b B^{\mu \nu}_{\lambda}(\vec{r}_{12})B^{\lambda \rho}_{\xi}(\vec{r}_{23}) }{(r_{12}\ r_{23})^d} \bar{\chi} {\cal D}^{DC} \gamma^\xi \partial_b \chi & -i\frac{f^{ACD}(r_{32})_b \,B^{\mu \rho}_{\lambda}(\vec{r}_{13})B^{\lambda \nu}_{\xi}(\vec{r}_{32})}{(r_{13} \ r_{32})^d} \bar{\chi} {\cal D}^{DB} \gamma^\xi \partial_b \chi & -i\frac{f^{BCD}(r_{13})_b B^{\nu \rho}_{\lambda}(\vec{r}_{23})B^{\lambda \mu}_{\xi}(\vec{r}_{13}) }{(r_{23} \ r_{31})^d}\bar{\chi} {\cal D}^{AD} \gamma^\xi \partial_b \chi\\
   \\
   &   i \frac{f^{ABD} \delta^{DC}B^{\mu \nu}_{\lambda}(\vec{r}_{12})C^{\lambda \rho}(\vec{r}_{23}) }{r^d_{12}\ r_{23}^{2d}} &  i\frac{f^{ACD}\delta^{DB}B^{\mu \rho}_{\lambda}(\vec{r}_{13})C^{\lambda \nu}(\vec{r}_{32})}{r^d_{13} \ r_{32}^{2d}} &  i\frac{f^{BCD}\delta^{DA} B^{\nu \rho}_{\lambda}(\vec{r}_{23})C^{\lambda \mu}(\vec{r}_{31}) }{r^{d}_{23} \ r_{31}^{2d}}\\
   \\
\end{bmatrix}}
\end{split}
\end{equation}
\begin{equation}
\begin{split}
\hat{\cal \psi} \equiv {\cal J}^{A,\mu}_{\psi} (\vec{r}_1){\cal J}^{B,\nu}_{\psi} (\vec{r}_2){\cal J}^{C,\rho}_{\psi} (\vec{r}_3) \qquad \qquad  \qquad \qquad \qquad \qquad \qquad \qquad  \\
\\
{\small \begin{bmatrix}
    &k_\psi \frac{\delta^{AB}C^{\mu \nu}(\vec{r}_{12})}{(r_{12})^{2d}}{\cal J}^{C,\rho}_{\psi} (\vec{r}_3)&k_\psi \frac{\delta^{AC}C^{\mu \rho}(\vec{r}_{13})}{(r_{13})^{2d}}{\cal J}^{B,\nu}_{\psi} (\vec{r}_2)&k_\psi \frac{\delta^{BC}C^{\nu \rho}(\vec{r}_{23})}{(r_{23})^{2d}}{\cal J}^{A,\mu}_{\psi} (\vec{r}_1) \\
    \\
   & -  \frac{f^{ABD} f^{DCE} B^{\mu \nu}_{\lambda}(\vec{r}_{12})B^{\lambda \rho}_{\xi}(\vec{r}_{23}) }{(r_{12}\ r_{23})^d} {\cal J}_{\psi}^{E,\xi}(\vec{r}_{3}) & - \frac{f^{ACD}f^{DBE}B^{\mu \rho}_{\lambda}(\vec{r}_{13})B^{\lambda \nu}_{\xi}(\vec{r}_{32})}{(r_{13} \ r_{32})^d} {\cal J}_{\psi}^{E,\xi}(\vec{r}_{3})& - \frac{f^{BCD}f^{DAE} B^{\nu \rho}_{\lambda}(\vec{r}_{23})B^{\lambda \mu}_{\xi}(\vec{r}_{31}) }{(r_{23} \ r_{31})^d}{\cal J}_{\psi}^{E,\xi}(\vec{r}_{3})\\
   \\
   &  -i\frac{f^{ABD}(r_{23})_b B^{\mu \nu}_{\lambda}(\vec{r}_{12})B^{\lambda \rho}_{\xi}(\vec{r}_{23}) }{(r_{12}\ r_{23})^d} \bar{\psi} {\cal D}^{DC} \gamma^\xi \partial_b \psi & -i\frac{f^{ACD}(r_{32})_b \,B^{\mu \rho}_{\lambda}(\vec{r}_{13})B^{\lambda \nu}_{\xi}(\vec{r}_{32})}{(r_{13} \ r_{32})^d} \bar{\psi} {\cal D}^{DB} \gamma^\xi \partial_b \psi & -i\frac{f^{BCD}(r_{13})_b B^{\nu \rho}_{\lambda}(\vec{r}_{23})B^{\lambda \mu}_{\xi}(\vec{r}_{13}) }{(r_{23} \ r_{31})^d}\bar{\psi} {\cal D}^{AD} \gamma^\xi \partial_b \psi\\
   \\
   &   i \frac{f^{ABD} \delta^{DC}B^{\mu \nu}_{\lambda}(\vec{r}_{12})C^{\lambda \rho}(\vec{r}_{23}) }{r^d_{12}\ r_{23}^{2d}} &  i\frac{f^{ACD}\delta^{DB}B^{\mu \rho}_{\lambda}(\vec{r}_{13})C^{\lambda \nu}(\vec{r}_{32})}{r^d_{13} \ r_{32}^{2d}} & i\frac{f^{BCD}\delta^{DA} B^{\nu \rho}_{\lambda}(\vec{r}_{23})C^{\lambda \mu}(\vec{r}_{31}) }{r^{d}_{23} \ r_{31}^{2d}}\\
   \\
\end{bmatrix}}
\end{split}
\end{equation}
\newline
Kronecker product between these two matrices provides us with all the possible terms. There is a subtlety involving the convention for the position of the resulting current operator. If we look at the first term in the second row ($\hat{\psi}_{21}$ or $\hat{\chi}_{21}$), in it we contract the currents at positions $\vec{r}_1$ and $\vec{r}_2$ and choose the resulting current to be at the position $\vec{r}_2$. We are free to make that choice, meaning we could have chosen the resulting current to be at $\vec{r}_1$ and it would not change the result. When we contract this resulting current in the final B contration with current at position $\vec{r}_3$ we have only one choice in terms of the structure factor indexes and signs of B functions since $|\vec{r}_2|<|\vec{r}_3|$ i.e. we have that the space-time point $\vec{r}_2$ is before $\vec{r}_3$. The same is true for the term $\hat{\psi}_{23}$ and $\hat{\chi}_{23}$. On the other hand in the expression in $\hat{\psi}_{22}$ that first B contraction will determine the result. We could choose the resulting current of that first contraction to be at $\vec{r}_1$ in which case it comes before $\vec{r_2}$ point with which the second B contraction is made. We could also choose it to be at $\vec{r}_3$ in which case it is after the point $\vec{r}_2$. These two different choices are at the heart of the reason why only terms that involve $\hat{\chi}_{22}$ and $\hat{\psi}_{22}$ will be the sole contributions to the BBC diagrams.
\newline
We have two types of contributions to the $Z_\lambda$. One is a result of the product of the elements of the first row in matrix $\hat{\psi}$ with the second row of matrix $\hat{\chi}$ and vice-versa. We call those terms BBC contributions since they have such products in the numerator. The other contribution is so-called CC contributions that comes from multiplying first row of matrix $\psi$ with the first row of matrix $\chi$. First we look at BBC contributions. Before that 

\subsection*{\centering Diagrams that have BBC in the numerator}
In third order and for $d=2$, we saw that only the "product" of the terms in the first row of $\hat{\cal \psi}$ and $\hat{\cal \chi}$ contributes and it is of order $k_\psi$ (and also the first row of $\hat{\cal \chi}$ with the second row of $\hat{\cal \psi}$ which is of order $k_{\chi}$ ). Both $k_\psi$ and $k_\chi$ terms have 9 distinct contractions. However, some terms out of those 9 are zero due to structure factors. That reduces the number of terms to six distinct ones:
\begin{equation}
    (\hat{\psi}_{11}\hat{\chi}_{22}+\hat{\psi}_{11}\hat{\chi}_{23})+(\hat{\psi}_{12}\hat{\chi}_{21}+ \hat{\psi}_{12}\hat{\chi}_{23})+(\hat{\psi}_{13}\hat{\chi}_{21}+ \hat{\psi}_{13}\hat{\chi}_{22})
\end{equation}
First we shall look at the $\hat{\psi}_{11}\hat{\chi}_{22}$ and $\hat{\psi}_{13}\hat{\chi}_{22}$ contributions. Explicitly, we look at the full expression:
\begin{equation}
   4\lambda^{3}(g_{d})^4 \int d^d \vec{r}_1\int d^d\vec{r}_2\int d^d\vec{r}_3 \  \Theta(r_{12},r_{13},r_{23}) \ (\hat{\psi}_{11}\hat{\chi}_{22}+\hat{\psi}_{13}\hat{\chi}_{22})
\end{equation}
Where $\Theta(r_{12},r_{13},r_{23}) = \theta(R>r_{12}>a)\theta(R>r_{13}>a)\theta(R>r_{23}>a)$. 
We can write each term explicitly as:
\begin{equation}
\begin{split}
    &\hat{\psi}_{11} \hat{\chi}_{22} = -k_\psi f^{ACD} f^{DBE} \delta^{AB} \ \frac{ B^{\mu \rho}_{\lambda}(\vec{r}_{13})B^{\lambda \nu}_{\xi}(\vec{r}_{12}) C_{\mu \nu}(\vec{r}_{12}) }{r_{13}^d\ r_{12}^{3d}}{\cal J}^{C,\rho}_{\psi} (\vec{r}_3) {\cal J}_{\chi}^{E,\xi}(\vec{r}_{3})\\
    &\hat{\psi}_{13} \hat{\chi}_{22} = -k_\psi f^{ACD} f^{BDE} \delta^{BC} \ \frac{ B^{\mu \rho}_{\lambda}(\vec{r}_{13})B^{\lambda \nu}_{\xi}(\vec{r}_{23}) C_{\nu \rho}(\vec{r}_{23}) }{r^d_{13}\ r_{23}^{3d}}{\cal J}^{A,\mu}_{\psi} (\vec{r}_1) {\cal J}_{\chi}^{E,\xi}(\vec{r}_{3})
\end{split}
\label{eq:surviving}
\end{equation}
A little comment is in order. In the above expressions we have made two different conventions for the second B contraction. In the first term ($\hat{\psi}_{11} \hat{\chi}_{22}$) we have chosen $\vec{r}_{2(13)} \to \vec{r}_{12}$ i.e. we have placed the result of the first B contraction at $\vec{r}_1$ which is to say before $\vec{r}_{2}$ and accordingly the structure factors are reflecting that. In the second term ($\hat{\psi}_{13} \hat{\chi}_{22}$) we have chosen $\vec{r}_{2(13)} \to \vec{r}_{23}$ i.e. we have chosen to place the result of the first B contraction after the $\vec{r}_2$. We are free to do so. From this we see that the two terms are the same.
\begin{equation}
\begin{split}
     -2k_\psi f^{ACD} f^{BDE} \delta^{BC} \ \frac{ B^{\mu \nu}_{\lambda}(\vec{r}_{13})B^{\lambda \rho}_{\xi}(\vec{r}_{23}) C_{\nu \rho}(\vec{r}_{23}) }{r^d_{13}\ r_{23}^{3d}}{\cal J}^{A,\mu}_{\psi} (\vec{r}_1) {\cal J}_{\chi}^{E,\xi}(\vec{r}_{3})
\end{split}
\end{equation}
Let us look at the second of the integrals because that one is more familiar. We have:
\begin{equation}
    A^{\mu}_{\xi} = B^{\mu \nu}_{\lambda}(\vec{r}_{13})B^{\lambda \rho}_{\xi}(\vec{r}_{23}) C_{\nu \rho}(\vec{r}_{23}) = 2r_{23}^2 (2(\vec{r}_{13}\cdot \vec{r}_{23})\delta^{\mu}_{\xi}+ d (r_{13})^\mu(r_{23})_\xi - 2 (r_{13})_{\xi} (r_{23})^\mu) 
\end{equation}
The only symmetric part of the above expressions is when $\mu = \xi$ so we have:
\begin{equation}
    A^{\mu}_{\mu} =  2r_{23}^2 (2(\vec{r}_{13}\cdot \vec{r}_{23})\delta^{\mu}_{\xi}+ d (r_{13})^\mu(r_{23})_\mu - 2 (r_{13})_{\mu} (r_{23})^\mu) \to 2\bigg(3-\frac{2}{d}\bigg)(\vec{r}_{23}\cdot \vec{r}_{23})(\vec{r}_{13}\cdot \vec{r}_{23})
\end{equation}
Where we have summed over all $\mu$'s in $A^{\mu}_{\mu}$ and $\mu = 1,...,d$.
This finally leads to the integral:
\begin{equation}
\begin{split}
    -2\lambda^3 (g_d)^4 k_\psi C_2 \, 2\bigg(3-\frac{2}{d}\bigg) \, \int d^d \vec{r}_1\int d^d\vec{r}_2\int d^d\vec{r}_3 \Theta \frac{(\vec{r}_{23}\cdot \vec{r}_{23})(\vec{r}_{13}\cdot \vec{r}_{23})}{r^d_{13}\ r_{23}^{3d}}{\cal J}^{A,\mu}_{\psi} (\vec{r}_1) {\cal J}_{\chi}^{E,\xi}(\vec{r}_{1})
\end{split}
\end{equation}
In the above integral we can change variables again, this time $\vec{r}_1 \leftrightarrow \vec{r}_3$ and in such a way we arrive at:
\begin{equation}
\begin{split}
    -2\lambda^3 (g_d)^4 k_\psi C_2 \, 2\bigg(3-\frac{2}{d}\bigg) \, \int d^d \vec{r}_1\int d^d\vec{r}_2\int d^d\vec{r}_3 \Theta \frac{(\vec{r}_{12}\cdot \vec{r}_{12})(\vec{r}_{13}\cdot \vec{r}_{12})}{r^d_{13}\ r_{12}^{3d}}{\cal J}^{A,\mu}_{\psi} (\vec{r}_1) {\cal J}_{\chi}^{E,\xi}(\vec{r}_{1})
\end{split}
\end{equation}
Just for the sake of checking the signs I want to write the other terms that contribute. Namely there is also this pair:
\begin{equation}
\begin{split}
    &\hat{\psi}_{12} \hat{\chi}_{21} = -k_\psi f^{ABD} f^{DCE} \delta^{AC} \ \frac{ B^{\mu \nu}_{\lambda}(\vec{r}_{12})B^{\lambda \rho}_{\xi}(\vec{r}_{23}) C_{\mu \rho}(\vec{r}_{13}) }{r_{12}^d\ r_{23}^{d} \ r_{13}^{2d}}{\cal J}^{B,\nu}_{\psi} (\vec{r}_2) {\cal J}_{\chi}^{E,\xi}(\vec{r}_{3})\\
    &\hat{\psi}_{13} \hat{\chi}_{21} = -k_\psi f^{ABD} f^{DCE} \delta^{BC} \ \frac{ B^{\mu \nu}_{\lambda}(\vec{r}_{12})B^{\lambda \rho}_{\xi}(\vec{r}_{23}) C_{\nu \rho}(\vec{r}_{23}) }{r_{12}^d\ r_{23}^{3d}}{\cal J}^{A,\mu}_{\psi} (\vec{r}_1) {\cal J}_{\chi}^{E,\xi}(\vec{r}_{3})
\end{split}
\end{equation}
The above terms diagramatically are similar to the diagram for $g_{(3,1)}$ in Fig.2 in the paper draft. The first ($\hat{\psi}_{12} \hat{\chi}_{21}$) has the double red contraction between (1) and (3) and the second one ($\hat{\psi}_{13} \hat{\chi}_{21}$) has the double red contraction between (2) and (3). Blue contraction goes from (1) to (2) and the result of that contracts with (3).
We also have two other terms:
\begin{equation}
\begin{split}
    &\hat{\psi}_{11} \hat{\chi}_{23} = -k_\psi f^{BCD} f^{DAE} \delta^{AB} \ \frac{ B^{ \nu \rho}_{\lambda}(\vec{r}_{23})B^{\lambda \mu}_{\xi}(\vec{r}_{13}) C_{\mu \nu}(\vec{r}_{12}) }{r_{23}^{d} \ r_{13}^{d} \ r_{12}^{2d}}{\cal J}^{C,\rho}_{\psi} (\vec{r}_3) {\cal J}_{\chi}^{E,\xi}(\vec{r}_{3})\\
    &\hat{\psi}_{12} \hat{\chi}_{23} = -k_\psi f^{BCD} f^{DAE} \delta^{AC} \ \frac{ B^{ \nu \rho}_{\lambda}(\vec{r}_{23})B^{\lambda \mu}_{\xi}(\vec{r}_{13}) C_{\mu \rho}(\vec{r}_{13}) }{r_{23}^d\ r_{13}^{3d}}{\cal J}^{A,\nu}_{\psi} (\vec{r}_1) {\cal J}_{\chi}^{E,\xi}(\vec{r}_{3})
\end{split}
\end{equation}
In the first term ($\hat{\psi}_{11} \hat{\chi}_{23}$) the double red line is between (1) and (2) and in the second one ($\hat{\psi}_{12} \hat{\chi}_{23}$) it is between (1) and (3). The blue line is like in the Fig. 2 for both of them. Interesting thing is that $\hat{\psi}_{12} \hat{\chi}_{21}$ will cancel with the $\hat{\psi}_{11} \hat{\chi}_{23}$ which is evident if in the latter one we do $\vec{r}_2 \leftrightarrow \vec{r}_3$:
\begin{equation}
\begin{split}
    \hat{\psi}_{11} \hat{\chi}_{23}|_{\vec{r}_2 \leftrightarrow \vec{r}_3} & = k_\psi C_2 \ \frac{ B^{ \nu \rho}_{\lambda}(\vec{r}_{23})B^{\lambda \mu}_{\xi}(\vec{r}_{12}) C_{\mu \nu}(\vec{r}_{13}) }{r_{23}^{d} \ r_{12}^{d} \ r_{13}^{2d}}{\cal J}^{E,\rho}_{\psi} (\vec{r}_3) {\cal J}_{\chi}^{E,\xi}(\vec{r}_{3})\\
    & (\vec{r}_1 \leftrightarrow \vec{r_3}) \to k_\psi C_2 \ \frac{ B^{ \nu \rho}_{\lambda}(\vec{r}_{12})B^{\lambda \mu}_{\xi}(\vec{r}_{23}) C_{\mu \nu}(\vec{r}_{13}) }{r_{23}^{d} \ r_{12}^{d} \ r_{13}^{2d}}{\cal J}^{E,\rho}_{\psi} (\vec{r}_3) {\cal J}_{\chi}^{E,\xi}(\vec{r}_{3})\\
    & (\rho \leftrightarrow \nu) \to k_\psi C_2 \ \frac{ B^{ \rho \nu}_{\lambda}(\vec{r}_{12})B^{\lambda \mu}_{\xi}(\vec{r}_{23}) C_{\mu \rho}(\vec{r}_{13}) }{r_{23}^{d} \ r_{12}^{d} \ r_{13}^{2d}}{\cal J}^{E,\nu}_{\psi} (\vec{r}_3) {\cal J}_{\chi}^{E,\xi}(\vec{r}_{3})\\
    & (\rho \leftrightarrow \mu) \to k_\psi C_2 \ \frac{ B^{ \mu \nu}_{\lambda}(\vec{r}_{12})B^{\lambda \rho}_{\xi}(\vec{r}_{23}) C_{\mu \rho}(\vec{r}_{13}) }{r_{23}^{d} \ r_{12}^{d} \ r_{13}^{2d}}{\cal J}^{E,\nu}_{\psi} (\vec{r}_3) {\cal J}_{\chi}^{E,\xi}(\vec{r}_{3}) = -\hat{\psi}_{12} \hat{\chi}_{21}
\end{split}
\end{equation}
Which makes evident that the cancelation happens and instead of 6 terms we have four. The other two can be, easily reduced to (careful because there is a minus sign i.e. $f^{ABD} f^{DCE} \delta^{BC}=-C_2 \delta^{AE}$ ):
\begin{equation}
\begin{split}
    \hat{\psi}_{13} \hat{\chi}_{21}|_{\vec{r}_1 \leftrightarrow \vec{r}_2} &= -k_\psi C_2 \ \frac{ B^{\mu \nu}_{\lambda}(\vec{r}_{12})B^{\lambda \rho}_{\xi}(\vec{r}_{13}) C_{\nu \rho}(\vec{r}_{13}) }{r_{12}^d\ r_{13}^{3d}}{\cal J}^{E,\mu}_{\psi} (\vec{r}_1) {\cal J}_{\chi}^{E,\xi}(\vec{r}_{3})\\
    & (\vec{r}_1 \leftrightarrow \vec{r}_3) \to -k_\psi C_2 \ \frac{ B^{\mu \nu}_{\lambda}(\vec{r}_{23})B^{\lambda \rho}_{\xi}(\vec{r}_{13}) C_{\nu \rho}(\vec{r}_{13}) }{r_{23}^d\ r_{13}^{3d}}{\cal J}^{E,\mu}_{\psi} (\vec{r}_1) {\cal J}_{\chi}^{E,\xi}(\vec{r}_{3})\\
    & (\mu \leftrightarrow \nu) \to -k_\psi C_2 \ \frac{ B^{\mu \nu}_{\lambda}(\vec{r}_{23})B^{\lambda \rho}_{\xi}(\vec{r}_{13}) C_{\mu \rho}(\vec{r}_{13}) }{r_{23}^d\ r_{13}^{3d}}{\cal J}^{E,\mu}_{\psi} (\vec{r}_1) {\cal J}_{\chi}^{E,\xi}(\vec{r}_{3})\\
    & (\rho \leftrightarrow \mu) \to -k_\psi C_2 \ \frac{ B^{\nu \rho}_{\lambda}(\vec{r}_{23})B^{\lambda \mu}_{\xi}(\vec{r}_{13}) C_{\mu \rho}(\vec{r}_{13}) }{r_{23}^d\ r_{13}^{3d}}{\cal J}^{E,\nu}_{\psi} (\vec{r}_1) {\cal J}_{\chi}^{E,\xi}(\vec{r}_{3}) = - \hat{\psi}_{12} \hat{\chi}_{23}
\end{split}
\end{equation}
Looks like these two cancel as well. Therefore the only surviving terms are the two terms in Eq.~(\ref{eq:surviving}). Diagramatically these two terms correspond to diagrams that have a blue line going from (1) to (3) and then the result of that (which we choose to place at (1) for the $\hat{\psi}_{11}\hat{\chi}_{22}$ and at (3) for $\hat{\psi}_{13}\hat{\chi}_{22}$) is conected to (2). As we have seen the two remaining terms are equal and lead to the integral:
\begin{equation}
\begin{split}
    -8\lambda^3 (g_d)^4 k_\psi C_2 \, 2\bigg(3-\frac{2}{d}\bigg) \, \int d^d \vec{r}_1\int d^d\vec{r}_2\int d^d\vec{r}_3 \Theta \frac{(\vec{r}_{23}\cdot \vec{r}_{23})(\vec{r}_{13}\cdot \vec{r}_{23})}{r^d_{13}\ r_{23}^{3d}}{\cal J}^{A,\mu}_{\psi} (\vec{r}_1) {\cal J}_{\chi}^{E,\xi}(\vec{r}_{1})
\end{split}
\end{equation}
We can introduce the relative coordinate system like we did in two-dimensional calculation:
\begin{equation} 
\begin{split}
    &\vec{r}_1=\frac{1}{\sqrt{3}}\vec{r}+\frac{1}{\sqrt{2}}\vec{v} +\frac{1}{\sqrt{6}}\vec{w}\\
    &\vec{r}_3=\frac{1}{\sqrt{3}}\vec{r}-\frac{1}{\sqrt{2}}\vec{v} +\frac{1}{\sqrt{6}}\vec{w}\\
    &\vec{r}_2=\frac{1}{\sqrt{3}}\vec{r} - \frac{2}{\sqrt{6}}\vec{w}\\
\end{split}
\qquad \longrightarrow \qquad 
\begin{split}
    &\vec{r}_1-\vec{r}_3 =\sqrt{2}\vec{v}\\
    &\vec{r}_1-\vec{r}_2 = \frac{1}{\sqrt{2}}\vec{v} + \sqrt{\frac{3}{2}}\vec{w}\\
    &\vec{r}_2-\vec{r}_3 = \frac{1}{\sqrt{2}}\vec{v} - \sqrt{\frac{3}{2}}\vec{w}
\end{split}
\end{equation}
After the substitution the above integral becomes:
\begin{equation}
\begin{split}
    -16\lambda^3 (g_d)^4 k_\psi C_2 \, \bigg(3-\frac{2}{d}\bigg) \, I_{3} \int d^d \vec{r} \, {\cal J}^{A,\mu}_{\psi} (\vec{r}) {\cal J}_{\chi}^{A,\mu}(\vec{r})
\end{split}
\end{equation}
\begin{equation}
\begin{split}
I_{3} = \int d^d\vec{v}\int d^d\vec{w} \frac{\bigg | \, \frac{1}{\sqrt{2}}\vec{v} -\sqrt{\frac{3}{2}}\vec{w} \, \bigg |^2 \, \bigg ( \sqrt{2}\vec{v} \cdot \big [\frac{1}{\sqrt{2}}\vec{v} -\sqrt{\frac{3}{2}}\vec{w} \big ] \bigg )  }{\sqrt{2}|\vec{v}|^d \, \bigg [\big ( \, \frac{1}{\sqrt{2}}\vec{v} -\sqrt{\frac{3}{2}}\vec{w} \, \big ) \cdot \big ( \, \frac{1}{\sqrt{2}}\vec{v} -\sqrt{\frac{3}{2}}\vec{w} \, \big )  \bigg ]^{3d/2}} \ \Theta(\sqrt{2}|\vec{v}|, \, |\frac{1}{\sqrt{2}}\vec{v} + \sqrt{\frac{3}{2}}\vec{w}|, \, |\frac{1}{\sqrt{2}}\vec{v} -\sqrt{\frac{3}{2}}\vec{w}|) \
\end{split}
\end{equation}
We can now rescale $\vec{v} \to \sqrt{2}\vec{v}$ and $\vec{w} \to \sqrt{\frac{2}{3}}\vec{w}$ to get:
\begin{equation}
\begin{split}
I_{3} 
& =  2 \bigg ( \frac{2}{3} \bigg )^{d/2}\int d^d\vec{v}\int d^d\vec{w} \frac{ \, \bigg ( \vec{v} \cdot \big [ \vec{v}-\vec{w} \big ] \bigg )  }{|\vec{v}|^d \, \bigg [\big ( \,  \vec{v}-\vec{w} \, \big ) \cdot \big ( \,  \vec{v}-\vec{w} \, \big )  \bigg ]^{3d/2-1}} \ \Theta(2|\vec{v}|, \, |\vec{v} + \vec{w}|, \, |\vec{w}-\vec{v}|)
\end{split}
\end{equation}
\newline
\subsubsection*{\centering Doing the BBC integrals}
There is an old way we did these integrals (its commented out) but I think the methods Yashar used to calculate CC diagrams are simpler and better suited so I am going to use it here. We start from:
\begin{equation}
\begin{split}
I_{3} =  -2 \bigg ( \frac{2}{3} \bigg )^{d/2}\int d^d\vec{v}\int d^d\vec{w} \frac{ \, \bigg ( \vec{v} \cdot \big [\vec{w}-\vec{v} \big ] \bigg )  }{|\vec{v}|^d \, \bigg [\big ( \, \vec{w} - \vec{v} \, \big ) \cdot \big ( \, \vec{w} - \vec{v} \, \big )  \bigg ]^{3d/2-1}} \ \Theta(2|\vec{v}|, \, |\vec{v} + \vec{w}|, \, |\vec{w}-\vec{v}|)
\end{split}
\end{equation}
and we shift $\vec{w} \to \vec{w}+\vec{v}$ which leads to a integral:
\begin{equation}
\begin{split}
I_{3} =  -2 \bigg ( \frac{2}{3} \bigg )^{d/2}\int d^d\vec{v}\int d^d\vec{w} \frac{ \, \vec{v} \cdot \vec{w} }{|\vec{v}|^d \, |\vec{w}|^{3d-2}} \ \Theta(2|\vec{v}|, \, |\vec{w}+2\vec{v}|, \, |\vec{w}|)
\end{split}
\end{equation}
And then in addition we rescale $\vec{v} \to (1/2)\vec{v}$ to get:
\begin{equation}
\begin{split}
I_{3} =  -\frac{1}{2^d} \bigg ( \frac{2}{3} \bigg )^{d/2}\int d^d\vec{v}\int d^d\vec{w} \frac{ \, \vec{v} \cdot \vec{w} }{|\vec{v}|^d \, |\vec{w}|^{3d-2}} \ \Theta(|\vec{v}|, \, |\vec{w}+\vec{v}|, \, |\vec{w}|)
\end{split}
\end{equation}
Now, when we do the derivative with respect to the cutoff we get:
\begin{equation}
-\partial_a I_{3} =  \frac{1}{2^d} \bigg ( \frac{2}{3} \bigg )^{d/2}\int d^d\vec{v}\int d^d\vec{w} \frac{ \, \vec{v} \cdot \vec{w} }{|\vec{v}|^d \, |\vec{w}|^{3d-2}} \ \Big[\delta(|\vec{v}|-a) \Theta(|\vec{w}+\vec{v}|, \, |\vec{w}|)+ \delta(|\vec{w}|-a) \Theta(|\vec{w}+\vec{v}|, \, |\vec{v}|)+ \delta(|\vec{w}+\vec{v}|-a) \Theta(|\vec{w}|, \, |\vec{v}|)\Big]
\end{equation}
First two intergrals above are the same just with relabeling $\vec{w} \to \vec{v}$ and vice-versa. 
First integral we can solve by introducing spherical coordinates for $\vec{v}$ and cylindrical for $\vec{w}$, as well as setting $w_z$ to be along the $\vec{v}$:
\begin{equation}
\begin{split}
& \frac{1}{2^{d-1}} \bigg ( \frac{2}{3} \bigg )^{d/2} \Omega_{d} \Omega_{d-1}\int dw_z dw_\perp \int dv \ w^{d-2} v^{d-1}\frac{ \, v\, w_z }{v^d \, w^{3d-2}} \ \delta(v-a) \tilde{\Theta}(w^2 + v^2 + 2 v \, w_z, \, w^2) \\
& = \frac{1}{2^{d-1}} \bigg ( \frac{2}{3} \bigg )^{d/2} a^{3-2d} \, \Omega_{d} \Omega_{d-1}\int dw_z dw_\perp  \ w_{\perp}^{d-2} \frac{ \, w_z }{w^{3d-2}} \  \tilde{\tilde{\Theta}}(w_{\perp}^2 + (w_z+1)^2, \, w_\perp^2+w_z^2)
\end{split}
\end{equation}
The other integral is obrained by switching back $\vec{w}\to \vec{w}-\vec{v}$ but this time we introduce spherical coordinates for $\vec{w}$ and cylindrical for $\vec{v}$:
\begin{equation}
\begin{split}
& \frac{1}{2^d} \bigg ( \frac{2}{3} \bigg )^{d/2}\int d^d\vec{v}\int d^d\vec{w} \frac{ \, \vec{v} \cdot (\vec{w}-\vec{v}) }{|\vec{v}|^d \, |\vec{w}-\vec{v}|^{3d-2}} \ \delta(w-a) \Theta(|\vec{w}-\vec{v}|, \, |\vec{v}|)\\
&\qquad = \frac{1}{2^d} \bigg ( \frac{2}{3} \bigg )^{d/2} \, a^{3-2d} \, \Omega_d \Omega_{d-1}\int dv_z \int dv_\perp \,v_{\perp}^{d-2}\frac{v_z-v^2 }{v^d \, (1+v^2-2v_z)^{3d/2-1}} \ \tilde{\tilde{\Theta}}(1+v^2-2v_z, \, v^2)\\
\end{split}
\end{equation}
So, to combine them together in order to do the numerical integration we have that the both integrals acan be combined into:
\begin{equation}
\begin{split}
I_3= -\frac{1}{2^{d+1}} \bigg ( \frac{2}{3} \bigg )^{d/2} a^{3-2d} \, \Omega_{d} \Omega_{d-1}\int dx \, dy \bigg (\ & \frac{ 2\, |y|^{d-2} \, x }{(x^2+y^2)^{3d/2-1}} \  \Theta[(x+1)^2+ y^2, \, x^2+y^2]  \\
& \qquad + \, \frac{|y|^{d-2} \, (x-x^2-y^2) }{(x^2+y^2)^{d/2} \, [(x-1)^2+y^2)]^{3d/2-1}} \Theta[(x-1)^2+y^2, \, x^2+y^2] \bigg )
\end{split}
\end{equation}
Where we have relabeld variables and the Heavisides are $\Theta[(x+1)^2+ y^2, \, x^2+y^2] = \Theta[1<(x+1)^2+ y^2 < \, (R/a)^2]\, \Theta[1<x^2+ y^2 < \, (R/a)^2]$ and $\Theta[(x-1)^2+ y^2, \, x^2+y^2] = \Theta[1<(x-1)^2+ y^2 < \, (R/a)^2]\, \Theta[1<x^2+ y^2 < \, (R/a)^2]$. The two integral ranges could be reduced to the same range by doing the substitution $x\to-x$ in either one of the integrals:
\begin{equation}
\begin{split}
I_3= -\frac{1}{2^{d+1}} \bigg ( \frac{2}{3} \bigg )^{d/2} a^{3-2d} \, \Omega_{d} \Omega_{d-1}\int dx \, dy \bigg (\ & \frac{ 2\, |y|^{d-2} \, x }{(x^2+y^2)^{3d/2-1}}  \\
& \qquad - \, \frac{|y|^{d-2} \, (x+x^2+y^2) }{(x^2+y^2)^{d/2} \, [(x+1)^2+y^2)]^{3d/2-1}}  \bigg )\  \Theta[(x+1)^2+ y^2, \, x^2+y^2]
\end{split}
\end{equation}
Therefore this contribution to the $y_\lambda$ is:
\begin{equation}
\begin{split}
    -16\tilde{\lambda}^2 \lambda (g_d)^4  C_2 \, \bigg(3-\frac{2}{d}\bigg) (k_{\psi}+k_{\chi}) \frac{1}{2^{d}} \bigg ( \frac{2}{3} \bigg )^{d/2} \, \Omega_{d} \Omega_{d-1}  \, {\cal{I}}_{CBB}(d) \, d\ell\, 
\end{split}
\end{equation}
Where ${\cal I}_{BBC}$ is a numerical solution of the above integral in the expression for $I_3$. What is interesting about it is that it is positive for d under $d_c \approx 2.7$ after which it becomes negative and we no longer have the conformal window.
\begin{equation}
\begin{split}
    y^{CBB}_{\lambda} = -16 \tilde{\lambda}^2  (k_{\psi}+k_{\chi}) g_3^{BBC}, \qquad g_3^{BBC}= (g_d)^4  C_2 \, \bigg(3-\frac{2}{d}\bigg) \frac{1}{2^{d}} \bigg ( \frac{2}{3} \bigg )^{d/2} \, \Omega_{d} \Omega_{d-1}  \, {\cal{I}}_{BBC}, \qquad y_{\lambda}^{CBB}(d) >0 \ \text{for} \ d<2.7
\end{split}
\end{equation}
\subsubsection*{\centering Diagrams that have CC Contractions in the numerator:}
Integrals that contribute to the so called CC contractions come from multiplying the first row of the $\hat{\psi}$ with the first row of $\hat{\chi}$. In total there are 9 terms and we can write them all in a simple way as :
\begin{equation}
    I^{\mu}_{\rho} = 6 \frac{C^{\mu \nu}(\vec{r}_{12})C_{\nu \rho}(\vec{r}_{23})}{(r_{12} r_{23})^{2d}}
\end{equation}
Let us look at the second of the above terms. We can easily find what the product of C's is:
\begin{equation}
    C^{\mu \nu}(\vec{x}) C_{\nu \rho} (\vec{y}) = (2 x^{\mu} x^{\nu} - \delta^{\mu \nu}x^2)(2 y_{\nu}y_{\rho} - \delta_{\rho \nu}y^2)
\end{equation}
We only keep the symmetric terms in the above expression, i.e. ones that are invariant under simultaneous $x^{\mu} \to -x^{\mu}, \ y^{\rho} \to-y^{\rho}$ which makes the integral $I^{\mu}_{\rho} \to I^{\mu}_{\mu}$ and the numerator becomes: $C^{\mu \mu} (\vec{x}) C^{\mu \mu} (\vec{y})$. After introducing the change of coordinates and rescaling as well as an additional shift that Yashar does in his notes and in its rescaled version goes as: $\vec{w} \to \vec{w}+\vec{v}$ (Jacobian is 1 so no need to worry about it) we arrive at an integral:
\begin{equation}
    I^{\mu}_{\mu} = \frac{6}{3^{d/2}} \int d^d \vec{v} \int d^d \vec{w}\frac{C^{\mu \mu}(2\vec{v})C^{\mu \mu}(\vec{w})}{(v w)^{2d}} \Theta (2v,2v+w,w)
\end{equation}
Integrand is completely separable and the functions $C^{\mu \mu} (\vec{x})$ are known and for, say $\mu=z$ it is:
\begin{equation}
    C^{zz} (\vec{x}) = 2(x_z)^2 - x^2 \to x_z^2 - x_{\perp}^2 \to  x^2 (2\cos^2(\theta)  -1)
\end{equation}
in cylindrical and spherical coordinates respectively. We can rescale $\vec{v} \to \frac{1}{2}\vec{v}$ and do the derivative:
\begin{equation}
\begin{split}
   \partial_a I^{\mu}_{\mu} (a) & = \frac{6}{3^{d/2}} \int d^d \vec{v} \int d^d \vec{w}\frac{C^{\mu \mu}(\vec{v})C^{\mu \mu}(\vec{w})}{(v w)^{2d}} \Big[\delta(v-a)\Theta (|\vec{v}+\vec{w}|,w)+ \delta(w-a) \Theta (v,|\vec{v}+\vec{w}|)+\delta(|\vec{v}+\vec{w}|-a)\Theta (v,w)\Big]
\end{split}
\end{equation}
Again, the first two integrals are the same and there are only two integrals we need to solve.
First integral can be solved:
\begin{equation}
\begin{split}
   & \frac{12}{3^{d/2}} \int d^d \vec{v} \int d^d \vec{w}\frac{C^{\mu \mu}(\vec{v})C^{\mu \mu}(\vec{w})\delta(v-a)}{(v w)^{2d}} \Theta (|\vec{v}+\vec{w}|,w) \\
   & = \frac{12}{3^{d/2}} a^{3-2d} (d/2-1)\Omega_d \Omega_{d-1}\int dw_z \int dw_\perp w_\perp^{d-2} \frac{w_z^2 -w_\perp^2}{(w_z^2+w_\perp^2)^d} \tilde{\tilde{\Theta}} (1+w^2+2w_z,w^2)
\end{split}
\end{equation}
Where we have used an integral property $\int d\Omega_d (2 \cos^2(\theta)-1) = (2/d-1)\Omega_d$ and we can see that the zero contribution of the CC diagrams is a peculiarity of $d=2$ and not for $d\neq2$ dimensions. Let us look at the other integral. In it we make a substitution $\vec{\omega}\to \vec{w}-\vec{v}$ which simplifies it to:
\begin{equation}
\begin{split}
   & \frac{6}{3^{d/2}} \int d^d \vec{v} \int d^d \vec{w}\frac{C^{\mu \mu}(\vec{v})C^{\mu \mu}(\vec{w}-\vec{v})\delta(w-a)}{(v |\vec{w}-\vec{v}|)^{2d}}\Theta (v,|\vec{w}-\vec{v}|)\\
   & = \frac{6}{3^{d/2}} a^{3-2d} \bigg( \frac{2}{d}-1 \bigg)\Omega_d \Omega_{d-1}\int dv \, v^{3-2d} \int d \theta_w \, v \, \sin^{d-2}(\theta_w) \frac{(w_z-v)^2-w_\perp^2}{(w_\perp^2+(w_z-v)^2)^d} \theta (v^2, \, \omega_\perp^2+(w_z+1)^2) \Bigg |_{w \to 1}
\end{split}
\end{equation}
We introduced cylindrical coordinates for $\vec{w}$ and cylindrical for $\vec{v}$ and we choose for $w_z$ along $\vec{v}$. Now we can introduce a substitution:
\begin{equation}
(w_z-v)^2 +w_\perp^2 = v^2+1-2v \cos(\theta_w), \qquad (w_z-v)^2-w_\perp^2 = v^2+1-2 v \cos(\theta_w) -2 \sin^2(\theta_w)    
\end{equation}
and then using: $v_z = v \cos(\theta_w)$ and $v_\perp = v \sin(\theta_w)$ we finally arrive at:
\begin{equation}
\begin{split}
   & \frac{6}{3^{d/2}} a^{3-2d} \bigg( \frac{2}{d}-1 \bigg)\Omega_d \Omega_{d-1} \int dv_z \int dv_\perp |v_\perp|^{d-2} \frac{(v_z-1)^2+v_\perp^2 - 2v_\perp^2/(v_\perp^2+v_z^2)}{(v_z^2+v_\perp^2)^d \, [(v_z-1)^2+v_\perp^2]^d} \theta(v^2, \, (v_z-1)^2+v_\perp^2)\\
\end{split}
\end{equation}
Combining these two results we arrive at:
\begin{equation}
\begin{split}
I_3= \frac{6}{3^{d/2}}  a^{3-2d} \, \Omega_{d} \Omega_{d-1}\int dx \, dy \bigg (\ & \frac{ |y|^{d-2} \, (x^2-y^2) }{(x^2+y^2)^{d}} \  \Theta[(x+1)^2+ y^2, \, x^2+y^2]  \\
& \qquad + \, \frac{|y|^{d-2}}{2}\frac{[(x+1)^2 +y^2]-2y^2/(y^2+x^2)}{(x^2+y^2)^{d} \, [(x+1)^2+y^2)]^{d}} \Theta[(x+1)^2+y^2, \, x^2+y^2] \bigg )
\end{split}
\end{equation}
It gives the following result:
\begin{equation}
\begin{split}
    6 k_{\psi}k_{\chi}\,\tilde{\lambda}^3 (2g_d)^4 (\frac{2}{d}-1) \frac{1}{3^{d/2}} \, \Omega_{d} \Omega_{d-1}  \, {\cal{I}}_{CC} (d) \, d\ell\, 
\end{split}
\end{equation}
Where ${\cal I}_{CC}(d)$ si the numerical result of the above double integral. Finally, we can write this by separating the $d$-dependent part as:
\begin{equation}
\begin{split}
    y_{\lambda}^{CC} = 6 k_{\psi}k_{\chi} \tilde{\lambda}^2\,g^{CC}_{3}(d), \qquad g^{CC}_{3}(d)=(2g_d)^4 (\frac{2}{d}-1) \frac{1}{3^{d/2}} \, \Omega_{d} \Omega_{d-1}  \, {\cal{I}}_{CC} (d), \qquad y_\lambda^{CC}>0
\end{split}
\end{equation}
\subsubsection*{\centering $\beta$-function}
Therefore the full $y_{\lambda}$ at third order is:
\begin{equation}
\begin{split}
    y^{(3)}_{\lambda} = -16 (k_{\psi}+k_{\chi})g^{BBC}_{3}(d,N) +6 k_{\psi}k_{\chi}\,g^{CC}_{3}(d)
\end{split}
\end{equation}
This means that our $Z_{\lambda}$, after including both the one and two loop contributions becomes:
\begin{equation}
    Z_{\lambda} = (1+y_\lambda d\ell), \qquad y_\lambda = g_2(N,d)\tilde{\lambda}-16(k_{\psi}+k_{\chi}) \tilde{\lambda}^2f^{BBC}_{3}(N, d) +6 k_{\psi}k_{\chi}\,\tilde{\lambda}^2 g^{CC}_{3}(d), \qquad  g_2(N,d) = (3-2/d)\frac{C_2}{\Omega_d}
\end{equation}
Together with:
\begin{equation}
\begin{split}
    y_{\psi} = \frac{k_{\chi}}{\Omega_d} \frac{(N_f^2-1)}{N_f} \tilde{\lambda}^2(a), \qquad Z_{\psi}  = 1+y_\psi \, d\ell >1, \quad\text{and}\quad
    y_{\chi} = \frac{k_{\psi}}{\Omega_d} \frac{(N_f^2-1)}{N_f} \tilde{\lambda}^2(a), \qquad Z_{\chi}  = 1+y_\chi \, d\ell >1
\end{split}
\end{equation}
We can find the beta function, defined as:
\begin{equation}
    \beta = \tilde{\lambda}(-\epsilon +y_\lambda-y_\psi-y_\chi)
\end{equation}
\begin{equation}
\begin{split}
    \tilde{\beta} & = -\epsilon \, \tilde{\lambda} +g_2(N,d)\tilde{\lambda}^2 -16 (k_{\psi}+k_{\chi}) \tilde{\lambda}^3g^{CBB}_{3}(N, d) +6 k_{\psi}k_{\chi}\,\tilde{\lambda}^3 g^{CC}_{3}(d) -(k_\chi+k_\psi) f_2 (N,d) \tilde{\lambda}^3\\
    \beta& = -\epsilon \, \lambda +g_2\lambda^2 -b\lambda^3, \qquad \lambda = \tilde{\lambda}/N\\
\end{split}
\end{equation}
And we have:
\begin{equation}
\begin{split}
    &f_2(d,N)= \frac{1}{\Omega_d} (1-N^{-2}), \qquad g_2=(3-2/d)\frac{1}{\Omega_d}, \qquad b=16(\kappa_\psi+\kappa_\chi)\big [ g^{}_3(d) + f_2(d,N) \big]-6\kappa_{\chi}\kappa_{\psi}g^{\prime}_3(d)
\end{split}
\end{equation}
\begin{equation}
\begin{split}
    &g_3=\frac{1}{N}f^{CBB}_{3}(d)= 16(g_d)^4 \, \bigg(3-\frac{2}{d}\bigg) \frac{1}{2^{d}} \bigg ( \frac{2}{3} \bigg )^{d/2} \, \Omega_{d} \Omega_{d-1}  \, {\cal{I}}_{CBB}\\
    &g_3^\prime = f^{CC}_3(d) =6(2g_d)^4 \bigg( \frac{2}{d}-1 \bigg ) \frac{1}{3^{d/2}} \, \Omega_{d} \Omega_{d-1}  \, {\cal{I}}_{CC} (d)
\end{split}
\end{equation}
Important thing to note is that $g_3$ and $g^{\prime}_{3}$ are of order $\sim 1/\Omega^2_d$ while $f_2$ is of order $1/\Omega_d$ which for $d \approx 2$ makes $f_2 >> g_3, g_3^\prime$ and hence the only relevant contribution at two loop actually comes from the one-loop calculated wave function renormalization.
\subsection{Mass Deformations of the beta function}
Here, we drop the overall pre-factor or sign.

\emph{CC part} -- The CC term comes from the integral 
\be
I=\int{d^dv d^dw}\theta(a<v<R)\theta(a<w<R)\theta(a<\abs{\vec v+\vec w}<R)\frac{g^2(m_\psi v)}{\Omega_d^2v^{2d}}\frac{g^2(m_\chi w)}{\Omega_d^2w^{2d}}\tilde F_1(\vec v)\tilde F_1(\vec w)
\ee
where $\tilde F_1(\vec r)=F_1(\vec r)-r^2n(ra)$ in the massive case. The massless part $F_1(\vec r)$ can be written in a number of ways:
\be
F_1(\vec r)=2r_z^2-\vec r^2, \qqquad F_1(\vec r)=r_z^2-r_\perp^2, \qqquad F_1(\vec r)=r^2(2\cos^2\theta-1)
\ee
in cartesian, cylindrical and spherical coordinates respectively. 

This CC contribution has three parts, including $\delta(v-a)$, $\delta(w-a)$ and $\delta(\abs{\vec v+\vec w}-a)$. The first part does not get any interesting mass corrections, because we are interested in the limit $m_\psi a\ll 1$ and $m_\chi=0$. The non-trivial corrections are second and third term, but we can also find $I_1$ from $I_2$ by $m_\psi\lr m_\chi$ substitution. Taking derivative and rescaling
\bea
a^{2d-3}\partial_a I_2&=&3^{-d/2}\int{d^dv d^dw}\theta(1<v<R/a)\delta(w-1)\theta(1<\abs{\vec v+\vec w}<R/a)\frac{g^2(am_\psi v)}{\Omega_d^2v^{2d}}\frac{g^2(am_\chi w)}{\Omega_d^2w^{2d}}\tilde F_1(\vec v)\tilde F_1(\vec w)\nonumber\qquad\\
&&\hspace{6cm}\times\{F_1(\vec v)-v^2n(vam_\psi)\}\{F_1(\vec w)-w^2n^2(wam_\chi)\}
\eea
The functions $g(x)$ cut the IR. We define two new IR cut-off if the masses are larger than inverse system size $1/R$
\be
R_\psi\equiv \min(R,\frac{1}{m_\psi}), \qqquad R_\chi\equiv \min(R,\frac{1}{m_\chi})
\ee
We use spherical coordinate for $w$ and cylindrical coordinate for $v$ and separating the gapless part, find
\be
a^{2d-3}\partial_a I_2=a^{2d-3}\partial_a I_2^0+
3^{-d/2}\frac{\Omega_d\Omega_{d-1}}{\Omega_d^4}(1-\frac{2}{d})\int v_\perp^{d-2} dv_\perp dv_z \frac{n(avm_\psi)}{v^{2d-2}}\theta(1<v<R_\psi/a)\theta(1<\sqrt{v_\perp^2+(v_z+1)^2}<R_\psi/a)\nonumber
\ee
At this point we define a new parameter $\vec x=m_\psi\vec va$ and using that $m_\psi a\ll1 $ find
\be
a^{2d-3}\partial_a I_2=a^{2d-3}\partial_a I_2^0+
3^{-d/2}\frac{\Omega_d\Omega_{d-1}}{\Omega_d^4}(1-\frac{2}{d})(m_\psi a)^{d-2}\int x_\perp^{d-2} dx_\perp dx_z \frac{n(x)}{x^{2d-2}}\theta\Big[0<x<{\rm min}(1,m_\psi R)\Big]
\ee
At this point we go to polar coordinate and define $x_\perp=x\cos\phi$
\bea
a^{2d-3}\partial_aI_1(a)&=&a^{2d-3}\partial_aI_1^0(a)-3^{-d/2}\frac{\Omega^v_d\Omega^w_{d-1}}{\Omega_d^4}\{\frac{2}{d}-1\}(m_\chi a)^{d-2}\int\abs{\cos\phi}^{d-2} d\phi \int_0^{{\rm min}(1,m_\psi R)}dx x^{1-d}
n^2(x)
\eea
We are interested in the regime where the mass is sufficiently large so that ${\rm min}(1,m_\psi R)=1$. In that case
\be
\int_0^{2\pi}\abs{\cos\phi}^{d-2}d\phi=2\sqrt{\pi}\frac{\Gamma(d/2-1)}{\Gamma(d/2)}
\andd \int_0^{1}dx x^{1-d}n^2(x)\approx \frac{1}{2}\label{eqIntx}\quad\text{for $d\in(2,3.5)$}.
\ee
Note that $\Gamma(d/2-1)$ is diverging and this cancels the $(1-2/d)$ prefactor. Finally we find
\bea
a^{2d-3}\partial_aI_2(a)
&=&a^{2d-3}\partial_aI_2^0(a)+3^{-d/2}\frac{\Omega_d\Omega_{d-1}}{\Omega_d^4}(m_\psi a)^{d-2}\frac{2\sqrt{\pi}}{d}
\eea
But note that if mass is smaller than $1/R$ so that $m_\psi R\ll 1$, one can substitute $n(x)\to x/(d-2)$, and the integral \pref{eqIntx} results in $(mR)^{4-d}/(d-2)^2$, leading to $\propto(ma)^{2}(a/R)^{d-4}\approx 0$ which vanishes. Likewise, the third term gives
\bea
a^{2d-3}\partial_a I_3=3^{-d/2}\int{d^dvd^dw}\theta(1<v<R/a)\theta(1<w<R/a)\delta(\abs{\vec v+\vec w}-1)\frac{g^2(am_\psi v)}{\Omega_d^2v^{2d}}\frac{g^2(am_\chi w)}{\Omega_d^2w^{2d}}\tilde F_1(\vec v)\tilde F_1(\vec w)
\eea
We do two change of variables, first $\vec w\to \vec w-\vec v$ and then $\vec v\to -\vec v$. This results in
\bea
a^{2d-3}\partial_a I_3&=&3^{-d/2}\int{d^dvd^dw}\theta(1<v<R/a)\theta(\abs{\vec v+\vec w}-1)\delta(w-1)\frac{g^2(am_\psi v)}{\Omega_d^2v^{2d}}\frac{g^2(am_\chi \abs{\vec w+\vec v})}{\Omega_d^2\abs{\vec w+\vec v}^{2d}}\\
&&\hspace{4cm}\{F_1(-\vec v)-v^2n^2(vam_\psi)\}\{F_1(\vec v+\vec w)-\abs{\vec v+\vec w}^2n^2(\abs{\vec v+\vec w}am_\chi)\}
\eea
Again, the $g(x)$ sets new IR cut-offs. Using spherical coordinate for $w$ and cylindrical coordinate for $v$ we have
\bea
a^{2d-3}\partial_a I_3&=&3^{-d/2}\frac{\Omega_d\Omega_{d-1}}{\Omega_d^4}\int{v_\perp^{d-2}dv_\perp dv_z}\frac{\theta(1<\sqrt{v_\perp^2+v_z^2}<R_\psi/a)}{(v_\perp^2+v_z^2)^{d}}\frac{\theta(1<\sqrt{v_\perp^2+(v_z+1)^2}<R_\chi/a)}{[v_\perp^2+(v_z+1)^2]^{d}}\\
&&\hspace{0cm}\{(v_z^2-v_\perp^2)-(v_z^2+v_\perp^2)n^2(\sqrt{v_z^2+v_\perp^2}am_\psi)\}\{[(v_z+1)^2-v_\perp^2]-[(v_z+1)^2+v_\perp^2]n^2(\sqrt{(v_z+1)^2+v_\perp^2}am_\chi)\}\nonumber
\eea
Defining $m=\max(m_\psi,m_\chi)$ and $\vec x=ma\vec v$ we have
\bea
a^{2d-3}\partial_a I_3&=&3^{-d/2}\frac{\Omega_d\Omega_{d-1}}{\Omega_d^4}(ma)^{3d-4}\int{x_\perp^{d-2}dx_\perp dx_z}\frac{\theta(ma<\sqrt{x_\perp^2+x_z^2}<mR_\psi)}{(x_\perp^2+x_z^2)^{d}}\frac{\theta(1<\sqrt{x_\perp^2+(x_z+ma)^2}<mR_\chi)}{[x_\perp^2+(x_z+ma)^2]^{d}}\nonumber\\
&&\hspace{-1.5cm}\{(x_z^2-x_\perp^2)-(x_z^2+x_\perp^2)n^2(\frac{m_\psi}{m}\sqrt{x_z^2+x_\perp^2})\}\{[(x_z+ma)^2-x_\perp^2]-[(x_z+ma)^2+x_\perp^2]n^2(\frac{m_\chi}{m}\sqrt{(x_z+ma)^2+x_\perp^2}am_\chi)\}\nonumber
\eea
At this point, the expression is symmetric in $m_\psi\lr m_\chi$. Now, assume $m_\chi\to 0$ and take the $m_\psi a\ll 1$ and $m R_\psi\to 1$ 
limit, to find
\bea
a^{2d-3}\partial_a I_3&=&a^{2d-3}\partial_a I_3^0-3^{-d/2}\frac{\Omega_d\Omega_{d-1}}{\Omega_d^4}(ma)^{3d-4}\int{x_\perp^{d-2}dx_\perp dx_z}\frac{\theta(0<\sqrt{x_\perp^2+x_z^2}<1)}{(x_\perp^2+x_z^2)^{2d}}(x_z^4-x_\perp^4)n^2(\sqrt{x_z^2+x_\perp^2})\qquad
\eea
where we have separated the gapless contribution. At this point, we can go to polar coordinate and compute the integral, but we won't bother, because in $d=2+\eps$, it is proportional to $(m_\psi a)^2$ and vanishes in $m_\psi a\ll 1$ limit.
\sepline
\emph{BBC part} -- For the (BB)C contraction 
\be
I^\mu_\alpha=\int d^2vd^2w g^3(m_\psi r_{12})g(m_\chi r_{23})\frac{A^\mu{}_\alpha(\vec r_{23},\vec r_{12})}{\Omega_d^3 r_{23}^d r_{12}^{3d}}\theta(a<r_{12}<R)\theta(a<r_{23}<R)\theta(a<r_{13}<R)
\ee
Defining $\vec x=\vec r_{12}$ and $\vec y=\vec r_{23}$, we have $\vec r_{13}=\vec x+\vec y$ and we need
\be
A^\mu{}_\alpha(\vec x,\vec y)=B^{\mu\nu}{}_\lambda (\vec x)B^{\lambda\rho}{}_\alpha(\vec y)C_{\rho\nu}(\vec y)=2y^2[2(\vec x\cdot\vec y)\delta^\mu_\alpha+x^\mu y_\alpha d-2x_\alpha y^\mu]
\ee
In presence of the mass, this is modified to
\bea
A^\mu{}_\alpha(\vec x,\vec y)&\to& A^\mu{}_\alpha(\vec x,\vec y)+y^2n^2(my)B^{\mu\nu}{}_\lambda (\vec x)[B^\lambda_{\nu}]_\alpha(\vec y)\\
&=& A^\mu{}_\alpha(\vec x,\vec y)+y^2n^2(my)[(2-d)x^\mu y_\alpha]
\eea
where we have used that
\be
B^{\mu\nu}{}_\lambda (\vec x)[B^\lambda_{\nu}]_\alpha(\vec y)=[x^\mu\delta^\nu_\lambda+x^\nu\delta^\mu_\lambda-x_\lambda\delta^{\mu\nu}][y^\lambda \delta_{\nu\alpha}+y_\nu\delta^\lambda_\alpha-\delta^\lambda_\nu y_\alpha]=(2-d)x^\mu y_\alpha
\ee
We focus on the diagonal component which means
\be
A^\mu{}_\alpha(\vec x,\vec y)\to y^2(\vec x\cdot\vec y)\delta^\mu_\alpha {\cal A}, \qqquad {\cal A}_0\equiv 2[2+(d-2)/d], \qqquad {\cal A}\equiv \Big\{{\cal A}_0-n^2(my)(d-2)/d\Big\}
\ee
As usual we do a change of variables to
\be
\vec r_{12}=\sqrt{2}\vec v \qquad \vec r_{23}=-\frac{\vec v}{\sqrt{2}}+\sqrt{\frac32}\vec w, \qquad \vec r_{13}=\frac{\vec v}{\sqrt 2}+\sqrt{\frac{3}{2}}\vec w
\ee
First the shift $\vec w\to \vec w+\vec v/\sqrt{3}$ and then re-scaling $\vec v\to \vec v/\sqrt{2}$ and $\vec w\to \sqrt{2/3}\vec w$ we have
\bea
I=3^{-d/2}\int{d^dvd^dw}g^3(m_\psi v)g(m_\chi w)\frac{A^\mu{}_\alpha(\vec w,\vec v)}{\Omega^d w^dv^{3d}} \theta_1(a<v<R)\theta_2(a<w<R)\theta_3(a<\vert{\vec v+\vec w}\vert<R)
\eea
First, we consider the gapless limit. Taking the derivative produces three terms:
\be
a^{2d-3}\partial_a I_1=\frac{\Omega_d\Omega_{d-1}}{\Omega_d^4}{\cal A}_0\int{dw_zdw_\perp}\frac{w_zw_\perp^{d-2}}{(w_z^2+w_\perp^2)^{d/2}}\theta(1<\sqrt{w_z^2+w_\perp^2})\theta(1<\sqrt{w_\perp^2+(w_z+1)^2})
\ee
\be
a^{2d-3}\partial_a I_2=\frac{\Omega_d\Omega_{d-1}}{\Omega_d^4}{\cal A}_0\int{dv_zdv_\perp}\frac{v_zv_\perp^{d-2}}{(v_z^2+v_\perp^2)^{3d/2-1}}\theta(1<\sqrt{v_z^2+v_\perp^2})\theta(1<\sqrt{v_\perp^2+(v_z+1)^2})
\ee
Lastly we can make the third term in the form of $A(\vec w+\vec v,-\vec v)/(\vec w+\vec v)^dv^{3d}$ multiplied by $\delta(w-1)\theta(v>1)\theta(\abs{w+v}>1)$:
\be
a^{2d-3}\partial_a I_3=\frac{\Omega_d\Omega_{d-1}}{\Omega_d^4}{\cal A}_0\int{dv_zdv_\perp}\frac{-v_zv_\perp^{d-2}[v_z+(v_z^2+v_\perp^2)]}{[v_\perp^2+(v_z+1)^2]^{d/2}(v_z^2+v_\perp^2)^{3d/2-1}}\theta(1<\sqrt{v_z^2+v_\perp^2})\theta(1<\sqrt{v_\perp^2+(v_z+1)^2})
\ee
Now, we consider the massive case. Note that $C^{\mu\nu}(\vec y)$ is the only one modified. Therefore, $I_1$ is not affected.
\bea
a^{2d-3}\partial_a I_2&\to& a^{2d-3}\partial_a I_2^0-\frac{d-2}{d}\int{dv_zdv_\perp}\frac{v_zv_\perp^{d-2}}{(v_z^2+v_\perp^2)^{3d/2-1}}n(m_\psi a\sqrt{v_z^2+v_\perp^2})\\
&&\hspace{4cm}\times \theta(1<\sqrt{v_z^2+v_\perp^2}<R_\psi/a)\theta(1<\sqrt{v_\perp^2+(v_z+1)^2}<R_\psi/a)\nonumber
\eea
which could be written in the form
\bea
a^{2d-3}\partial_a I_2&\to& a^{2d-3}\partial_a I_2^0-\frac{d-2}{d}(ma)^{2d-3}\int{d x_zd x_\perp}\frac{ x_z x_\perp^{d-2}}{( x_z^2+ x_\perp^2)^{3d/2-1}}n(\sqrt{x_z^2+x_\perp^2})\\
&&\hspace{4cm}\times\theta(m_\psi a<\sqrt{x_z^2+x_\perp^2}<m_\psi R_\psi)\theta(m_\psi a<\sqrt{x_\perp^2+(x_z+m_\psi a)^2}<m_\psi R_\psi)\nonumber
\eea
And again we are interested in the limit when $m_\psi R_\psi=1$. It may appear that in $d=2+\eps$ this is linear in $(ma)$. However, if we take the $m_\psi a\to 0$ in the integrand, the integral vanishes due to odd-ness in $x_z$. Therefore the integral is at least $\propto (ma)$, and the full result $\propto (ma)^2$ or higher power.

Likewise, $I_3$ is also modified
\bea
a^{2d-3}\partial_a I_3&=&a^{2d-3}\partial_a I_3^0-\frac{d-2}{d}\int{dv_zdv_\perp}\frac{-v_zv_\perp^{d-2}[v_z+(v_z^2+v_\perp^2)]}{[v_\perp^2+(v_z+1)^2]^{d/2}(v_z^2+v_\perp^2)^{3d/2-1}}n(m_\psi a\sqrt{v_z^2+v_\perp^2})\\
&&\hspace{6cm}\theta(1<\sqrt{v_z^2+v_\perp^2}<R_\psi/a)\theta(1<\sqrt{v_\perp^2+(v_z+1)^2<R_\psi/a})\nonumber
\eea
which becomes
\bea
a^{2d-3}\partial_a I_3&=&a^{2d-3}\partial_a I_3^0+\frac{d-2}{d}(ma)^{3d-5}\int{d x_zd x_\perp}\frac{ x_z x_\perp^{d-2}[ma x_z+( x_z^2+ x_\perp^2)]}{( x_z^2+ x_\perp^2)^{4d/2-1}}n(\sqrt{ x_z^2+ x_\perp^2})\\
&&\hspace{5cm}\theta(m_\psi a<\sqrt{ x_z^2+ x_\perp^2}<m_\psi R_\psi)\theta(m_\psi a<\sqrt{ (x_z+m_\psi a)^2+ x_\perp^2}<m_\psi R_\psi)\nonumber
\eea

\section{\large\sc Fierz Identity}
Suppose we have $N_c^2$ matrices defined so that
\be
\tr{\lambda^A\lambda^B}=\delta^{AB}, \quad A=0\dots N^2-1, \qquad \lambda^0=\frac{\bb 1_f}{\sqrt{N}}
\ee
In addition to $N^2-1$ generators, we have defined properly normalized $\lambda^0$. Then, the resolution of the identity is
\be
\sum_{A=0}^{N^2-1}(\lambda^A)_{fg}(\lambda^A)_{hi}=\delta_{fi}\delta_{gh}, \orr
\sum_{A=1}^{N^2-1}(\lambda^A)_{fg}(\lambda^A)_{hi}=\delta_{fi}\delta_{gh}-\frac{1}{N}\delta_{fg}\delta_{hi}\label{eqFI}
\ee
Multiplying it by $M_{fg}=\bar\psi_f\gamma^\mu\psi_g$ and $N_{hi}=\bar\chi_h\gamma^\nu\chi_i$ and summing over repeated flavor index we have
\bea
\sum_{A=1}^{N^2-1}(\bar\psi\lambda^A\gamma^\mu\psi)(\bar\chi\lambda^A\gamma^\nu\chi)
&=&(\bar\psi_f\gamma^\mu\psi_g)(\bar\chi_g\gamma^\nu\chi_f)-\frac{1}{N}(\bar\psi\gamma^\mu\psi)(\bar\chi\gamma^\nu\chi)
\eea

For four-component spinors in 2+1D one has the spinor Fierz identity, which is built from the 16  matrices $\Lambda^{\mu\nu}=\sigma^\mu\tau^\nu\sim \gamma^\mu \tau^\nu$. Define these matrices in the following way, the set of 16 hermitian 4$\times$4 matrices are
\be 
\Lambda^A\in\Big\{\bb 1, \quad\gamma^\mu,\quad \tau^\nu,\quad \gamma^\mu \tau^\nu\Big\} \so \tr{\Lambda^A\Lambda^B}=4\delta^{AB}
\ee
and these generate a Fierz identity:
\be
\delta_{fi}\delta_{gh}=\frac{1}{4}\sum_{A=0}^{15}(\Lambda^A)_{fg}(\Lambda^A)_{hi} 
\ee
Contracting this with $M=\bar\psi_{f\alpha}(\gamma^\nu\chi)_{g\alpha}$ and $N=\bar\chi_{h\beta}(\gamma^\nu\psi)_{i\beta}$ gives
\bea
&&-4(\bar\psi_{\alpha}\gamma^\nu\psi_{\beta})(\bar\chi_\beta\gamma^{\nu'}\chi_\alpha)=\sum_A(\bar\psi_\alpha \Lambda^A\gamma^\nu\chi_\alpha)(\bar\chi_\beta \Lambda^A\gamma^{\nu'}\psi_\beta)=(\bar\psi\gamma^\nu\chi)(\bar\chi\gamma^{\nu'}\psi)+(\bar\psi\gamma^\mu\gamma^\nu\chi)(\bar\chi\gamma^\mu\gamma^{\nu'}\psi)\\
&&\hspace{8cm}+(\bar\psi\vec \tau\gamma^\nu\chi)(\bar\chi\vec \tau\gamma^{\nu'}\psi)+(\bar\psi\gamma^\mu \vec \tau\gamma^\nu\chi)(\bar\chi\gamma^\mu \vec \tau\gamma^{\nu'}\psi)
\eea
Contracting $\nu$ and $\nu'$ and using that $\gamma^\mu\gamma^\nu=\delta^{\mu\nu}\bb 1+i\eps^{\mu\nu\lambda}\gamma^\lambda$, 
\bea
&&-4(\bar\psi_{\alpha}\gamma^\nu\psi_{\beta})(\bar\chi_\beta\gamma^{\nu}\chi_\alpha)=(\bar\psi\chi)(\bar\chi\psi)+(\bar\psi\vec\tau\chi)\cdot(\bar\chi\vec\tau\psi)\label{eqSFI}
\eea
Combining Eqs.\,\pref{eqSFI} and \pref{eqFI} gives the expression quoted in the paper. In the 1+1D case, dropping one of the $\gamma$ matrices, but still keeping $\{\gamma^\mu,\gamma^\nu\}=2\delta^{\mu\nu}\bb 1_{4\times 4}$ can be treated similarly. It leads to
\be
-2\sum_{\mu=0,1}(\bar\psi_\alpha\gamma^\mu\psi_\beta)(\bar\chi_\beta\gamma^\mu\chi)=\Big[(\bar\psi\chi)(\bar\chi\psi)+(\bar\psi\vec\tau\chi)\cdot(\bar\chi\vec\tau\psi)\Big]-\Big[(\bar\psi\gamma^5\chi)(\bar\chi\gamma^5\psi)+(\bar\psi\gamma^5\vec\tau\chi)\cdot(\bar\chi\gamma^5\vec\tau\psi)\Big]
\ee
where $\gamma^5=i\gamma^0\gamma^1$. Since the $\gamma^5$ interactions are in the repulsive channel, they are not important for the dynamical mass generation.
\section{\large\sc The O(4) symmetry}
Here, we investigate the energetics of various patterns of hybridization in the psuospin-valley (PSV) space. The Lagrangian is
\be
{\cal L}=\mat{\bar\psi & \bar\chi}\mat{\cancel k+m & V \\ V\dg & \cancel k}\mat{\psi\\\chi}
\ee
We are going to assume that $\gamma^\mu=\sigma^\mu\tau^0$ is purely in $\sigma$ and not in the $\tau$ or PSV space. But $m$ and $V$ are matrices acting in the $\tau$ space. For simplicity, we assume $V=V\dg$ is hermitian. Using $\bar\psi=\psi\dg\gamma^0$ and $\gamma^0=\sigma^z$ the Hamiltonian is
\be
{\cal H}_{8\times 8}=\mat{\vec k\cdot\vec\sigma+m\sigma^z & \sigma^zV \\ \sigma^z V & \vec k\cdot\vec\sigma}
\ee
Let me introduce another set of Pauli-matrices $\vec\mu$ acting in the species ($\psi$ or $\chi$ space). Then,
\be
{\cal H}=\vec k\cdot\vec\sigma\mu^0+ \sigma^zV\mu^x+m\sigma^z\frac{1+\mu^z}{2}
\ee
We work out the spectrum:
\be
{\cal H}^2=k^2+V^2+\frac{1}{2}m^2(1+\mu^z)+\frac{1}{2}\Big(\{V,m\}\mu^x-i[V,m]\mu^y\Big)
\ee
Now, suppose $m=m_0\hat m$ is along one of the four $\tau^\nu$ directions, and $V=V_0\hat V$ is in an arbitrary direction. It either commutes or anti-commutes (not both) and therefore, the magnitude of the last term in paranthesis is the same. 
\begin{table}[h!]
\begin{tabular}{c||c|c|c|c}
$m/V$ & $\tau^0$ & $\tau^x$ & $\tau^y$ & $\tau^z$\\
\hline\hline
$\tau^0$ & $2\mu^x$ & $2\mu^x\tau^x$ & $2\mu^x\tau^y$ & $2\mu^x\tau^z$\\
\hline
$\tau^x$ & $2\mu^x\tau^x$ & $2\mu^x$ & $-2\mu^y\tau^z$ & $2\mu^y\tau^y$\\
\hline
$\tau^y$ & $2\mu^x\tau^y$ & $2\mu^y\tau^z$ & $2\mu^x$ & $-2\mu^y\tau^x$\\
\hline
$\tau^z$ & $2\mu^x\tau^z$ & $-2\mu^y\tau^y$ & $2\mu^y\tau^y$ & $2\mu^x$\\
\end{tabular}
\end{table}

Therefore, we never have both $\mu^x$ and $\mu^y$ in the last term. Squaring twice we have
\be
\Big[E_k^2-(k^2+V_0^2+{m_0^2}/{2})\Big]^2=m_0^4/4+V_0^2m_0^2
\ee
So, we see that the spectrum in all 4$\times$4 cases is exactly the same.

\ew

\end{document}